\documentclass[11pt,a4paper]{article}

\usepackage{jheppub}
\usepackage[utf8]{inputenc}
\usepackage{amsmath}
\usepackage{amsthm}
\usepackage{amssymb}
\usepackage{enumitem}   
\usepackage{url}
\usepackage{mathtools}
\usepackage{slashed}
\usepackage{multirow}
\usepackage{xcolor}
\usepackage{slashed}
\usepackage{multirow}
\usepackage[toc,page]{appendix}
\usepackage{hyperref}

\usepackage{amsmath, amsfonts, amssymb}
\usepackage{graphicx}
\usepackage{color}
\usepackage{enumerate}
\usepackage{hyperref}
\usepackage{latexsym}
\usepackage{enumerate}
\usepackage{soul}
\usepackage[normalem]{ulem}
\usepackage{wasysym}
\usepackage{makecell}
\usepackage{slashed}
\usepackage{comment,verbatim}
\usepackage{amsmath}
\usepackage{subfig}

\newcommand{\SC}{{\rm sc}}
\newcommand{\fv}{{\rm fv}}

\newcommand{\DW}{{\rm DW}}
\newcommand{\GeV}{{\rm GeV}}
\newcommand{\nuc}{{\rm n}}
\newcommand{\perc}{{\rm p}}

\newcommand{\Z}{{\mathbb{Z}}}

\newcommand{\be}{\begin{equation}}
\newcommand{\ee}{\end{equation}}

\begin{document}



\title{
Electroweak Phase Transition with a Double Well Done Doubly Well
}
\author[a]{Prateek Agrawal,}
\author[b,c]{Simone Blasi,}
\author[c]{Alberto Mariotti,}
\author[a, d]{and Michael Nee}

\affiliation[a]{Rudolf Peierls Centre for Theoretical Physics,
University of Oxford, Parks Road, Oxford OX1 3PU, United Kingdom}
\affiliation[b]{Deutsches Elektronen-Synchrotron DESY, Notkestr.~85, 22607 Hamburg, Germany}
\affiliation[c]{Theoretische Natuurkunde and IIHE/ELEM, Vrije Universiteit
Brussel, \& The  International Solvay Institutes, Pleinlaan 2, B-1050 Brussels, Belgium}
\affiliation[d]{Department of Physics, Harvard University, Cambridge, MA, 02138, USA}
\emailAdd{prateek.agrawal@physics.ox.ac.uk}
\emailAdd{simone.blasi@desy.de}
\emailAdd{alberto.mariotti@vub.be}
\emailAdd{mnee@fas.harvard.edu}


\abstract{
We revisit the electroweak phase transition in the scalar
singlet extension of the standard model with a $\mathbb{Z}_2$ symmetry. 
In significant parts of the parameter space the phase transition occurs in two steps -- including canonical benchmarks used in experimental projections for gravitational waves. 
Domain walls produced in the first step of the transition seed the final step to the electroweak vacuum, an effect which is typically neglected but leads to an exponentially enhanced tunnelling rate. 
We improve previous results obtained for the seeded transition, which made use of the thin-wall or high temperature approximations, by using the mountain pass algorithm that was recently proposed as a useful tool for seeded processes. 
We then determine the predictions of the seeded transition for the latent heat, bubble size and characteristic time scale of the transition. Differences compared to homogeneous transitions are most pronounced when there are relatively few domain walls per hubble patch, potentially leading to an enhanced gravitational wave signal. We also provide a derivation of the percolation criteria for a generic seeded transition, which applies to
the domain wall seeds we consider as well as to strings and monopoles.
}

\preprint{
\begin{flushright}
DESY-23-208
\end{flushright}
}

\maketitle

\section{Introduction}
\label{sec:introduction}

As the highest energy scale probed directly in
experiments, the electroweak scale is at the edge of our knowledge of high energy
physics.
Further motivated by naturalness arguments, new physics hiding around this scale is therefore a compelling possibility and
represents an important theoretical target, signals of which may be observed in next generation colliders.

Interestingly, the early universe can provide a complementary
probe of this new physics. In particular, the LISA observatory~\cite{LISA:2017pwj} has sensitivity at the right frequency to detect gravitational waves (GWs) produced during a strongly first order electroweak phase transition (EWPT) ~\cite{Caprini:2015zlo, Caprini:2019egz}.
While detailed studies on the lattice have shown that electroweak symmetry breaking at finite temperature is a crossover
within the standard model (SM) alone~\cite{Kajantie:1995kf,
Kajantie:1996mn,Karsch:1996yh, Aoki:1996pn, Gurtler:1997hr,
Laine:1998jb}, many extensions including new physics at the
electroweak scale predict a first order transition (see,
e.g.~\cite{Pietroni:1992in, Carena:1996wj, Delepine:1996vn,
Apreda:2001us, Huber:2015znp}). 
A  first order EWPT may explain the observed baryon asymmetry of
the universe~\cite{Kuzmin:1985mm, Shaposhnikov:1987tw, Rubakov:1996vz} and can produce an
observable GW signal~\cite{Kosowsky:1991ua,
Kosowsky:1992rz, Kosowsky:1992vn, Kamionkowski:1993fg}.\footnote{See ref.~\cite{Athron:2023xlk} for a review.} This has
motivated in depth studies of these kinds of models in order to
connect possible signals at LISA with collider
signatures such as deviations in the SM Higgs
couplings~\cite{Hashino:2016rvx, Huang:2016cjm, Hashino:2016xoj, Artymowski:2016tme, Beniwal:2017eik}.

Most studies to date have focused on electroweak phase transitions
occurring in homogeneous space time. These are characterised by a
uniform nucleation probability owing to random thermal and quantum
fluctuations of the fields~\cite{Coleman:1977py, Callan:1977pt,
Linde:1981zj}. However, this picture is modified if impurities are
present at the time of the transition~\cite{Steinhardt:1981mm,
Steinhardt:1981ec, Hosotani:1982ii, Jensen:1982jv,Witten:1984rs,Yajnik:1986tg,
Yajnik:1986wq, Hiscock:1987hn, Berezin:1987ea, Arnold:1989cq,
Berezin:1990qs, Preskill:1992ck, Kusenko:1997hj, Dasgupta:1997kn,
Metaxas:2000qf, Kumar:2008jb, Kumar:2009pr, Kumar:2010mv,
Pearce:2012jp, Lee:2013zca, Gregory:2013hja, Burda:2015isa, Mukaida:2017bgd, Canko:2017ebb, Kohri:2017ybt, Oshita:2018ptr,Jinno:2021ury,
Shkerin:2021zbf, Shkerin:2021rhy, Agrawal:2022hnf, Blasi:2022woz, Briaud:2022few, Jinno:2023vnr}. Around the
impurities the barrier between the false and the true vacuum can be
naturally smaller, and the nucleation probability is consequently enhanced.
This inhomogeneous, or seeded, transition is typically exponentially
faster than the homogeneous one (which can still occur
away from the impurities), and therefore dominates the phenomenology of
the phase transition.

A particularly useful and minimal setup to study these effects is the
SM extended with a real scalar singlet (xSM) with $\Z_2$ symmetry,
$S$~\cite{McDonald:1993ex, Burgess:2000yq, Espinosa:2007qk,
Profumo:2007wc, Barger:2007im, Espinosa:2008kw, Espinosa:2011ax,
Cline:2012hg, Profumo:2014opa, Curtin:2014jma, Feng:2014vea,
Craig:2014lda, Curtin:2016urg, Huang:2016cjm, Vaskonen:2016yiu, Kurup:2017dzf,
Buttazzo:2018qqp, Alanne:2019bsm,Carena:2019une,Schicho:2021gca,Cline:2021iff,Laurent:2022jrs,Azatov:2022tii}. In this model, a strong first order
EWPT is typically realised in two steps\footnote{See also\,\cite{Niemi:2020hto}
for a lattice study of a two--step electroweak phase transition and \cite{Gould:2023ovu} for an approach based on dimensional reduction.}.
 In the first step the $\Z_2$
symmetry is spontaneously broken via a non--zero vacuum expectation
value (vev) of $S$ while electroweak symmetry remains unbroken.
The breaking of the $\Z_2$ symmetry leads to the formation of domain
walls via the Kibble-Zurek mechanism~\cite{Zeldovich:1974uw,
Kibble:1976sj, Zurek:1985qw}. This phase is only metastable,
however, and in the second and final step of the transition the Higgs
acquires its electroweak vacuum expectation value (vev), $v_h$, and the $\Z_2$ symmetry is restored. This second step can be strongly first order and represents a
promising scenario for obtaining a detectable signal in GWs.

For parameter choices in the xSM which predict a first order EWPT, the model is often difficult to probe at the LHC. A $\Z_2$ symmetry of sufficient quality prevents higgs-singlet mixing and for a singlet mass $m_S$ which satisfies $m_S>m_h/2$ there are no corrections to higgs decays~\cite{Curtin:2014jma}. Bounds from phase transition dynamics and future GW experiments, in particular LISA, offer the only ways to probe the model in large regions of parameter space~\cite{Huang:2016cjm,Kurup:2017dzf}, although the model generically predicts modified couplings which could be observed at future colliders~\cite{Katz:2014bha}. Given the importance of this model for studies of electroweak baryogenesis~\cite{Profumo:2007wc, Cline:2012hg, Profumo:2014opa, Vaskonen:2016yiu} and dark matter~\cite{McDonald:1993ex, Burgess:2000yq}, an accurate determination of the phase transition dynamics is important in order to reliably test these models.

\begin{figure}
	\centering
	\includegraphics[scale=.28]{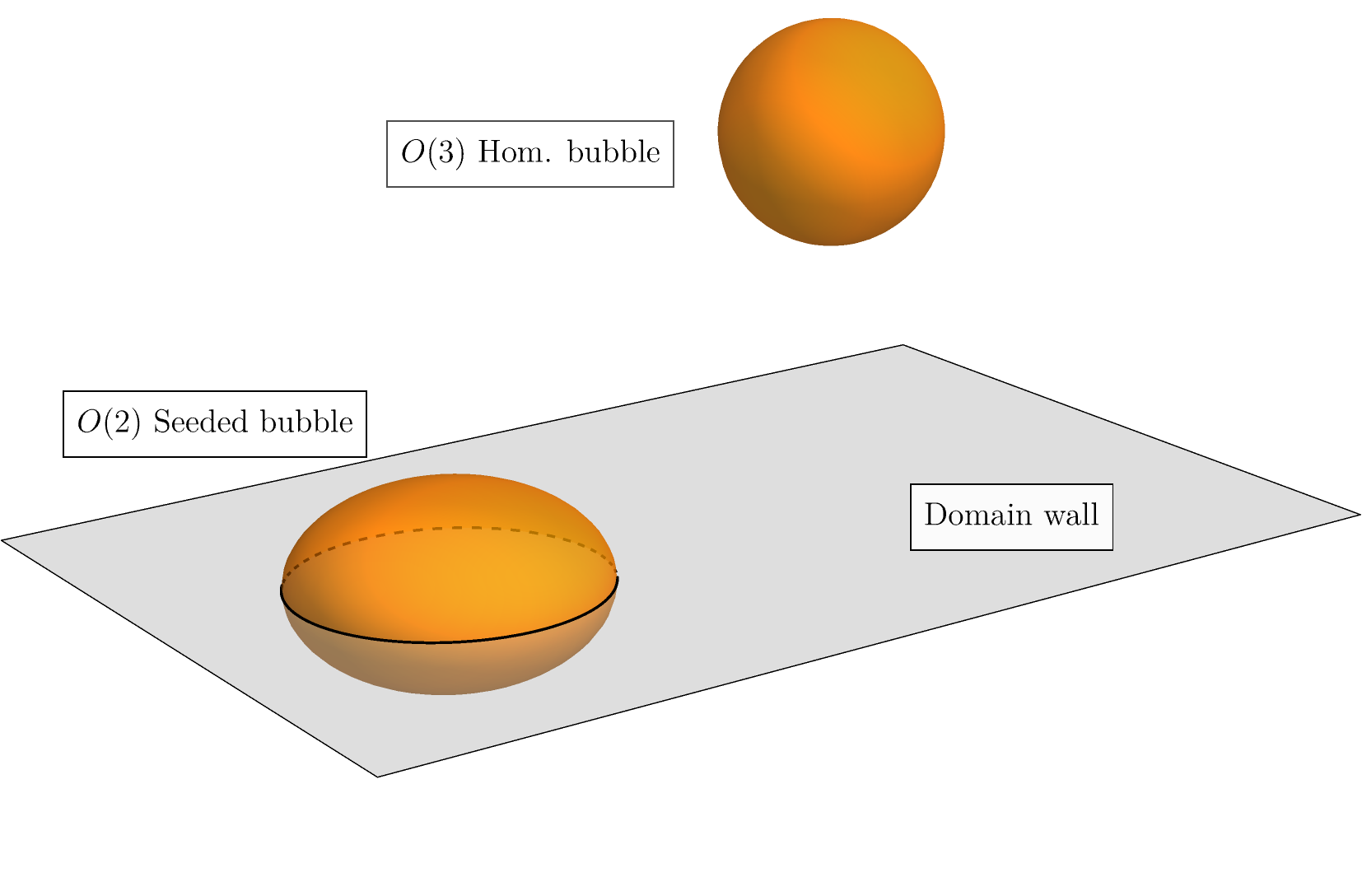}
	\caption{Cartoon of the seeded and homogeneous bubbles. Elliptical bubbles with an $O(2)$ symmetry are nucleated on the domain walls, while spherical bubbles with $O(3)$ symmetry are nucleated in the homogeneous spacetime far from the domain walls.}
\label{fig:cartoon}
\end{figure}

\smallskip
In this paper we continue the study of the seeded EWPT in the xSM initiated in ref.~\cite{Blasi:2022woz}. The catalysing impurities are the domain walls produced after the $\Z_2$ symmetry breaking in the first step of the transition. If any explicit $\Z_2$ breaking terms are small (as required for instance if $S$ is to be a dark matter candidate), then the domain walls will survive until the second step of the transition.
The domain walls then catalyze the formation of Higgs bubbles nucleated on the domain wall plane. These bubbles are not spherically symmetric due to the presence of the wall, but are elliptical with a reduced $O(2)$ symmetry, see figure~\ref{fig:cartoon} for a cartoon of the seeded and homogeneous bubbles.
Ref.~\cite{Blasi:2022woz} showed that the seeded transition is generically faster than the homogeneous one, and that regions of parameter space which are naively ruled out (because the homogeneous one is suppressed) can become viable due to the presence of domain walls acting as catalyzing seeds. 

The analysis in ref.~\cite{Blasi:2022woz} was carried out in the high--temperature limit. In addition, the seeded tunnelling probability was evaluated either in the thin wall limit, or within the lower dimensional theory on the domain wall after integrating out the Kaluza--Klein states along the orthogonal direction. While these methods provide a new qualitative picture of the seeded tunnelling, in certain temperature ranges neither of these approximations can be applied, leaving a gap in calculability. This prevents an accurate determination of thermodynamical quantities such as the latent heat and the nucleation rate. 

In this paper we overcome several limitations of the study in ref.~\cite{Blasi:2022woz} and provide a state-of-the-art analysis of seeded vacuum decay including the full one-loop thermal potential. The use of the mountain pass algorithm, first presented in ref.~\cite{Agrawal:2022hnf} for the case of monopole catalysed tunnelling, allows us to numerically solve the equations of motion in the presence of a domain wall background without resorting to an approximation scheme such as the high temperature expansion or the thin wall limit\,\footnote{See appendix~\ref{sec:highT} for a comparison with the previous results of ref.~\cite{Blasi:2022woz}.}. With these results we can determine the regions of parameter space where the catalysed phase transition nucleates while the homogeneous transition is too slow to complete. Even for parameters where the homogeneous transition is cosmologically fast, we confirm that the catalysed transition is the dominant process, being exponentially enhanced relative to the homogeneous decay.

\smallskip

A crucial quantity determining the phenomenology of a first order phase transition is its duration or time scale, usually indicated by the dimensionless quantity $\beta/H$, with $H$ the Hubble rate. For a homogeneous transition this is controlled by the gradient of the bounce action around the nucleation temperature.
On the other hand, a catalysed transition possesses two characteristic time scales: the first one is given by the inverse of the nucleation rate around the impurity, and the second one relates to the average distance between the impurities.
In the following, we shall indicate by \emph{sparse} a network in which the latter time scale is much greater than the one from the intrinsic nucleation rate, whereas the network will be referred to as \emph{dense} in the opposite regime. 
In our analysis we will highlight the differences between both networks and the homogeneous transition, with the most significant distinction being the de-correlation between the latent heat and the average bubble size for sparse networks. 
We will additionally derive
the conditions for successful percolation for decays seeded by impurities of arbitrary codimension $D$ by generalizing the results of ref.\,\cite{Guth:1981uk} for point--like defects such as monopoles.

The GW spectrum from catalysed transitions is analysed for both sparse and dense networks by using a thin wall approximation to the bounce action. This allows for an efficient scan of the relevant parameter space, and we leave a more complete analysis for future work. Generically, we find that when both processes are viable, the catalysed decay results in a smaller signal amplitude as bubbles nucleate closer to the critical temperature and less energy is released in the transition. However, the presence of defects allows the transition to complete in new regions of parameter space which can lead to a large latent heat being released.
For the case of sparse networks the bubble size at collision is also significantly larger, leading to the production of long--lasting sound waves. 
This results in an enhanced GW signal which is peaked at lower frequencies than is typical of homogeneous transitions occurring at the same temperature. This behaviour has been observed in simulations of the catalysed transition from domain walls~\cite{Blasi:2023rqi}, allowing us to connect our analysis with the expected GW signal.

\smallskip

The rest of this paper is organised as follows. In section~\ref{sec:setup} we discuss the model parameter space and the cosmological history of the model, focusing on the formation of domain walls and the details of the electroweak phase transition for parameters where it is first order. The details of the mountain pass algorithm and a comparison between the homogeneous and seeded phase transition is presented in section~\ref{sec:MPT}. In section~\ref{sec:PTparams} we discuss the impact of catalysing defects on the various thermodynamic quantities relevant for describing the phenomenology of the transition. Finally, in section~\ref{sec:gravwaves} we compare the features of the GW signal between the homogeneous and catalysed transitions.

\section{The xSM and its thermal history}
\label{sec:setup}

In this section we review the singlet-extended SM (xSM) and its thermal history in the
$\mathbb{Z}_2$ symmetric limit,
with an emphasis on domain wall formation and its consequences. 
The
tree--level scalar potential dictated by the $\mathbb{Z}_2$ symmetry under which $S$ is odd and all other fields are even is
\begin{equation}\label{eq:V4d}
 V(h,S) = - \mu_h^2 |\mathcal{H}|^2 + 
 \lambda |\mathcal{H}|^4 -\frac{\mu_s^2}{2} S^2
 +\frac{\eta}{4} S^4 + \kappa |\mathcal{H}|^2 S^2,
\end{equation}
where the gauge is chosen such that the Higgs vev has the form $\langle \mathcal{H} \rangle= (0,h)/\sqrt{2}$. In the regime of interest all parameters ($\eta, \lambda, \kappa,\mu_h^2, \mu_s^2$) are positive. We introduce the following counterterms in the potential,
\be 
	V_\text{ct} = \mu^2_{h, \text{ct}} |\mathcal{H}|^2 + \lambda_\text{ct} |\mathcal{H}|^4 
	+ \mu^2_{s, \text{ct}} S^2 + \eta_\text{ct} S^4,
\ee
which are fixed by the requirement that at $T=0$ the Higgs mass, the Higgs vev and the singlet mass are unchanged from their tree--level values at the global minimum
$(h,S)=(v_h,0)$, and that the vev of $S$ in the false vacuum $(h,S)=(0,v_s)$ is the same as at tree level. The renormalisation scale used in the one--loop Coleman--Weinberg potential~\cite{Coleman:1973jx},
\begin{align}
	V_{\rm CW} (h, S)  &= 
	\sum_{i = h, S} \frac{m_i^4 (h, S)}{64\pi^2} 
	\left( \log \left(\frac{m_i^2 (h, S)}{ \mu_0^2} \right) -\frac{3}{2} \right)
\end{align}
 is taken to be $\mu_0 = v_h = 246\,\text{GeV}$.

In this paper we include thermal corrections within the truncated full dressing scheme (see~\cite{Curtin:2016urg} for other possible schemes and related uncertainties). This amounts to substituting thermal masses obtained at leading order in the high temperature expansion in the one--loop effective potential,
\be 
m^2_i \rightarrow m^2_i + \Pi_i(T^2).
\label{eq:thermalsub}
\ee
where $m_i$ is the field-dependent mass for particle species $i = h, S$, and
\be
\begin{aligned}
	&  \Pi_h(T^2)=    c_h T^2
	\qquad \, \,
	&&c_h = \frac{2m_W^2 + m_Z^2 +m_h^2 + 2m_t^2}{4v_h^2} + \frac{\kappa}{12}
	\\
	&  \Pi_S(T^2) =  c_s T^2
	&&c_s = \frac{4\kappa + 3\eta}{12}
	\label{eq:T2coefficients}
\end{aligned}
\ee 
Adopting this scheme, the 1-loop thermal corrections to the potential are given by:
\begin{align}
	V_T (h, S; T) &= \sum_{i = h, S} \frac{T^4}{2\pi^2} 
	J_B \left( \frac{m_i^2(h, S) + \Pi_i(T^2)}{T^2 } \right) \, ,
	\\
	J_B(y^2) &= \int_0^\infty dx \, x^2 \log \left[ 1 -\exp \left( - \sqrt{x^2-y^2} \right)\right]
	\, .
\end{align}
The full thermal potential we use in our calculations is therefore:
\begin{align}
\label{eq:effectiveV}
	 V_\text{eff}(h,S;T) &= V(h, S) + V_{\rm CW} (h, S)  + V_T (h, S; T) 
\end{align}
after subtracting counterterms and making the substitution~\eqref{eq:thermalsub} in the thermal integral. 

In some of the discussion which follows and for our results in appendix~\ref{sec:highT} we will keep only the leading temperature corrections. In this limit the effective potential is
\begin{align}
	V_\text{high-T}(h,S;T) &=
	\frac{c_hT^2 - \mu_h^2}{2} h^2 + 
 	\frac{\lambda}{4} h^4 + \frac{c_sT^2 - \mu_s^2}{2} S^2
 	+\frac{\eta}{4} S^4 + \frac{\kappa}{2} h^2 S^2
	\, .
	 \label{eq:VhT}
\end{align}
Keeping only the leading temperature corrections is not in general a valid approximation, but simplifies the discussion and allows some of the calculations to be done analytically.
\\

\begin{figure}
	\centering
	\includegraphics[scale=.35]{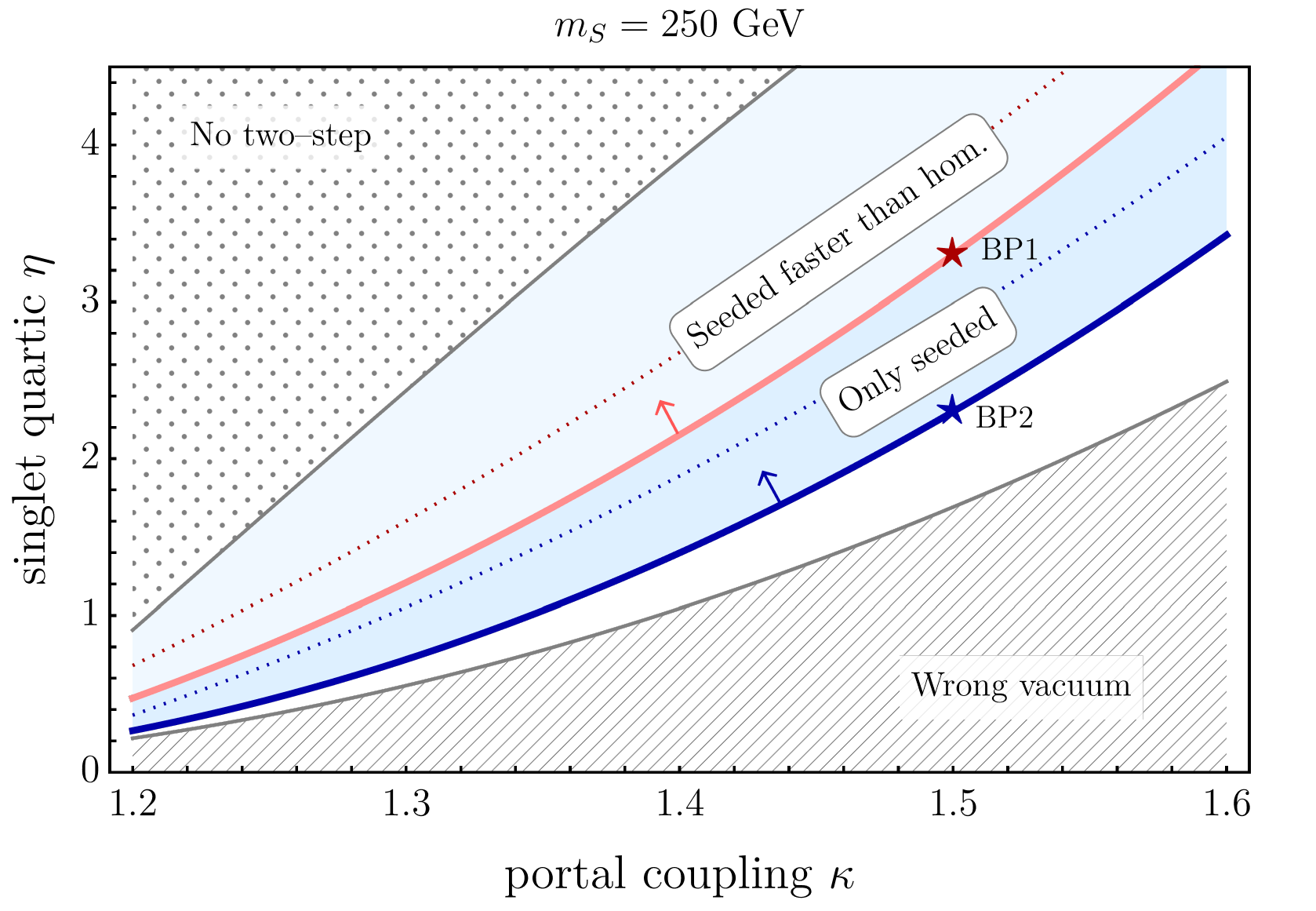}
	\caption{Parameter space for the electroweak phase transition in the xSM with $\Z_2$ symmetry. The region between the solid gray lines indicates where a two--step phase transition takes place. In the upper left corner the electroweak phase transition is second order, while the lower right corner is unable to reproduce the correct electroweak vacuum at zero temperature. Above the red solid line the homogeneous transition is cosmologically fast, but the catalyzed transition nucleates at higher temperatures. Between the red and blue solid lines only the seeded transition can lead to successful nucleation, while below the blue line the universe remains trapped in the false vacuum. Dotted lines have the same meaning of the solid lines but are obtained within the high--temperature approximation instead of the full one--loop thermal potential.}
\label{fig:scan}
\end{figure}

With the effective thermal potential $V_\text{eff}(h,S;T)$ at hand we can study the thermal history of the xSM. After fixing the higgs mass and vev to the observed values, the model is described by the three free parameters: $\{\kappa,\eta,m_S\}$, where $m_S$ is the singlet mass in the electroweak vacuum (at $T=0$). The regime we are interested in is the one where the electroweak phase transition is two step:
\be
\label{eq:twostep}
	(h, S) = (0,0) \rightarrow (0, \pm v_s) \rightarrow (v_h, 0)\, ,
\ee
which occurs for parameters such that 
\begin{align}
	\frac{\mu_s^2}{c_s} = \frac{\kappa v_h^2 - m_S^2}{c_s} \gtrsim \frac{\mu_h^2}{c_h} \, .
\end{align}
Fixing $m_S$ to representative values, the two--step region appears then as a stripe in the two dimensional plane parametrised by the portal coupling $\kappa$ and the singlet self-quartic $\eta$. This parameter space is shown in figure~\ref{fig:scan} between the two grey solid lines for singlet mass $m_S = 250$~GeV. In this region, the second step of the phase transition that breaks electroweak symmetry, $(0,\pm v_s) \rightarrow (v_h,0)$, is typically first order and the fate of the false vacuum depends on whether the tunnelling process is fast compared to the Hubble rate. 

The parameter space in which successful nucleation takes place has been identified in previous works by considering the homogeneous vacuum decay process, namely the nucleation of bubbles of true vacuum deep inside either the $+v_s$ or $-v_s$ domains. We show the parameter space where this process is cosmologically fast by the lighter blue region in figure~\ref{fig:scan}.  As we can see, there is a large part of the two--step parameter space where homogeneous nucleation is too slow and the Universe remains trapped in the false vacuum.

However, this neglects the effect of the defects formed at domain boundaries in the first step of the transition. In addition to the usual homogeneous false vacuum decay, the second step of the transition can be catalyzed by the domain walls that are formed in the first step~\cite{Blasi:2022woz}. The existence of this additional decay channel enlarges the viable parameter space where the phase transition can complete, which is shown by the darker blue region in figure~\ref{fig:scan}. Additionally, in the parameter space where both the homogeneous and seeded processes are possible, namely in the lighter blue region, the seeded process is always faster,\footnote{This assumes at least one domain wall per Hubble volume, see section~\ref{sec:catalyzednucl}.} and therefore dictates how the transition actually proceeds. 
 
The region where only seeded nucleation is cosmologically fast is identified in this work by taking into account thermal corrections beyond the leading high--$T$ terms, and represents one of the main results of our paper. The details of our calculation are presented in section~\ref{sec:catalyzednucl} as well as in section~\ref{sec:MPT} where we discuss our numerical strategy with reference to specific benchmark points BP1 and BP2 shown in figure~\ref{fig:scan}. 
 
\subsection{Formation of domain walls}

Due to the spontaneous $\Z_2$ symmetry breaking in the first step of Eq.\,\eqref{eq:twostep}, domain walls with a non trivial $S$ profile will form according to the Kibble mechanism. These are solutions of the classical equations of motion where $S$ interpolates between the two different vacua $\pm v_s$ related by the $\Z_2$. In the high temperature limit, a planar domain wall solution can be obtained exactly considering the potential in Eq.\,\eqref{eq:VhT}:
\begin{align}
	&S_{\DW}(z) =  v_S(T)
	\tanh \left(\frac{ \eta^{1/2} v_S(T) z}{\sqrt{2}} \right)
	\, ,
	&& v_S(T) =  \sqrt{ \frac{\mu_s^2 - c_s T^2}{\eta}} \,,
	\label{eq:DWprofile}
\end{align}
where $z$ is the coordinate orthogonal to the wall.

Away from the high--$T$ limit, the domain wall shape needs to be determined numerically as the solution to
\be
\label{eq:DW}
S^{\prime \prime}(z) = \frac{\partial}{\partial S} V_\text{eff}(0,S(z);T), \quad S(\pm \infty) = \pm v_s(T),
\ee
where the effective potential is given in Eq.\eqref{eq:effectiveV} and we have taken $h = 0$ according to the false vacuum configuration $(h, S) = (0,v_s(T))$.

The solution to \eqref{eq:DW} can be efficiently obtained by using the following first integral:
\be 
I = -\frac{1}{2} S^\prime(z)^2 + V_\text{eff}(0,S;T) = V_\text{eff}(0,v_s;T)
\ee
which fixes the slope of the domain wall profile at the center, $z=0$, as
\be
S^\prime(0) = \sqrt{2(V_\text{eff}(0,0;T) - V_\text{eff}(0,v_s;T))},
\ee 
where we have used the fact that $S(0)=0$ as the domain wall is odd for $z \rightarrow - z$. The soliton solution can then be simply obtained with the help of \texttt{Mathematica} by solving \eqref{eq:DW} with the above initial conditions for $S(0)$ and $S^\prime(0)$ without the need for shooting or relaxation algorithms.

The domain wall tension is given by
\be
\sigma_{\rm DW} = \int_{-\infty}^{\infty} {\rm d}z S^{\prime 2}(z) = \int_{-v_s}^{v_s} {\rm d}S \sqrt{2\left[V_{\rm eff}(0,S;T) - V_{\rm eff}(0,v_s;T)\right]}.
\ee
In general one has $\sigma_{\rm DW} \sim v_s(T)^3$, and in the simplest $\phi^4$ theory $\sigma_{\rm DW} = 2 \sqrt{2 \eta} v_s^3/3$.

\medskip

If the $\Z_2$ symmetry is exact, the domain walls are absolutely stable and will therefore survive until the second step of the phase transition takes place. However, if we allow for explicit $\Z_2$ breaking terms in the theory, the $S= \pm v_s$ vacua will no longer be degenerate. This introduces a pressure on the walls that eventually annihilates the domains with higher potential energy, leaving behind a homogeneous field configuration corresponding to the true vacuum~\cite{Larsson:1996sp, Casini:2001ai}. Considering as an example the case in which the $\Z_2$ breaking comes from a Planck-suppressed dimension-5 operator $c S^5 /M_P$, the domain walls will annihilate at a temperature\,\cite{Blasi:2022woz}
\begin{align}
	T_{\rm ann} \sim \frac{v_s}{2} 
	\frac{c^{1/2}}{\xi^{1/2} \eta^{1/4}}
	\, , 
\end{align}
where $\xi$ is the mean separation between domain walls (in Hubble units). We then see that for $\Z_2$ breaking due to gravity the symmetry can generically be of sufficient quality for domain walls to be long-lived enough to be present at the electroweak phase transition. In this work we consider the case of exact $\mathbb{Z}_2$ symmetry, neglecting any possible effects of small breaking terms. 

\subsection{Homogeneous false vacuum decay}
\label{sec:farfromwalls}

When the temperature drops below the critical temperature $T_c$, the $(v_h,0)$ vacuum becomes the global minimum and tunnelling can take place from the false vacuum $(0,v_s)$. Far from the domain walls, the nucleation rate can be found according to (euclidean) time independent and spherically symmetric solutions to the equations of motion~\cite{Coleman:1977py, Callan:1977pt, Linde:1981zj},
\be 
\label{eq:hombubble}
\phi^{\prime \prime}(r) + \frac{2}{r} \phi^\prime(r) = \frac{\partial}{\partial \phi} V_\text{eff}(h,S;T), 
\ee
with $\phi=h, S$, and boundary conditions 
\begin{align}
	S(\infty) = |v_s(T)|, \quad
	h(\infty) = 0,  \quad
	S^\prime(0) = 0, \quad
	h^\prime(0)=0 .
	\label{eq:hombcs}
\end{align}
Here $r$ indicates the radial coordinate of the spherical bubble. The equations \eqref{eq:hombubble}~\&~\eqref{eq:hombcs} can be readily solved with the numerical methods implemented in \texttt{FindBounce}~\cite{Guada:2020xnz} and \texttt{CosmoTransitions}~\cite{Wainwright:2011kj}. After determining the tunnelling field profiles, the nucleation rate per unit volume in the homogeneous false vacuum is given by
\begin{align}
	\Gamma_{\rm hom} \simeq c \,T^4 e^{-B_{\rm hom}(T)} \, ,
	\label{eq:homrate}
\end{align}
where $B_{\rm hom}(T)$ is the difference between the euclidean action evaluated on the field profiles satisfying the equations of motion~\eqref{eq:hombubble}
and the false vacuum solution
\begin{align}
	B_{\rm hom}(T) &= \frac{4\pi}{T} \int r^2 dr \left[ \mathcal{L}_E [h, S]
	- \mathcal{L}_E [0, v_s(T)] \right] \, .
	\label{eq:hombounce}
\end{align}
In the following we shall set $c = 1$ for concreteness. This is justified as varying $c$ causes a logarithmic correction to the dominant suppression factor, $B_{\rm hom}$, and so is a subleading effect.

For the homogeneous phase transition, the average number of bubbles at a temperature $T$ is given by
\begin{align}
	\mathcal{N}_{\, \rm hom} (T)= \int_{T}^{T_c} \frac{ d T'}{T'}  \,
	 \frac{\Gamma_{\rm hom}(T') }{ H(T')^{4}}\, ,
	\label{eq:nucleationhom}
\end{align}
with $\Gamma_{\rm hom}$ given in Eq.\,\eqref{eq:homrate}. This allows us to define the nucleation temperature, $T_\nuc$, as the temperature such that there is one bubble per hubble patch, $\mathcal{N}_{\, \rm hom} (T_\nuc) = 1$. For the present case of a phase transition at the electroweak scale, one finds $B_{\hom} \simeq 140$ as the nucleation condition. A different value for $c$ would shift this condition by $\sim \rm{log}\,c$.

\subsection{Catalysed false vacuum decay}
\label{sec:catalyzednucl}

The presence of domain walls introduces an additional channel for vacuum decay. Bubbles of true vacuum can form centered on the domain wall with a significantly enhanced probability compared to tunnelling inside the homogeneous false vacuum~\cite{Blasi:2022woz}. This process is governed by field profiles $h(\rho,z)$ and $S(\rho,z)$ which satisfy 
\be 
\label{eq:eomgen}
\frac{\partial^2 \phi}{\partial \rho^2} + \frac{1}{\rho} \frac{\partial \phi}{\partial \rho} + \frac{\partial^2 \phi}{\partial z^2} = \frac{\partial}{\partial \phi} V_\text{eff}(h,S;T), 
\ee 
with $\phi = h, S$ and boundary conditions
\be 
h(\infty,z) = h(\rho ,\pm \infty)= 0, \quad S(\infty,z) = S_\text{DW}(z), \,S(\rho, \pm \infty) = \pm v_s,
\label{eq:DWbcs}
\ee
where $S_\text{DW}(z)$ is the domain wall configuration obtained by solving \eqref{eq:DW}. Here $\rho = \sqrt{x^2 + y^2}$ is the radial coordinate on the domain wall plane, and $z$ indicates the orthogonal direction.
A typical solution to \eqref{eq:eomgen} satisfying \eqref{eq:DWbcs} is shown in figure~\ref{fig:profiles}.

As opposed to the system of equations describing the homogeneous transition (Eq.\,\eqref{eq:hombubble}), the profiles of the seeded bubbles do not respect the $O(3)$ rotational symmetry due to the presence of the wall in the $x-y$ plane, reducing the symmetry of the problem to only $O(2)$. This fact considerably complicates the task of finding the saddle configuration for the tunnelling process. The way we tackle this issue will be explained in section~\ref{sec:MPT}. 

\medskip

The bounce action for this tunnelling process is given by the euclidean action evaluated on the solutions to equations~\eqref{eq:eomgen}~\& \eqref{eq:DWbcs},
minus the action evaluated on the domain wall solution,
\begin{align}
	B_{\rm DW}(T) &= \frac{2\pi}{T} \int \rho d\rho dz
	 \left[ \mathcal{L}_E [h, S]
	- \mathcal{L}_E [0, S_\DW(z)] \right] \, .
	\label{eq:DWbounce}
\end{align}
The nucleation rate \emph{per unit surface} for bubbles nucleated on domain walls is given by
\begin{align}
	\Gamma_\DW(t) \simeq \sigma_{\rm DW} e^{-B_{\DW}(t)}
	\label{eq:DWrate}
\end{align}
where we estimate the prefactor to be set by the domain wall tension, $\sigma_{\rm DW} \sim v_s^3$. 

\medskip

For a seeded phase transition, the probability of nucleating a bubble per hubble volume depends on the nucleation rate $\Gamma_\text{DW}$ in eq.\,\eqref{eq:DWrate} as well as on the average number of domain walls inside the Hubble patch, $\xi$. Typically $\xi \sim \mathcal{O}(1)$ if the domain wall network has reached the scaling regime, but can be much larger in the immediate aftermath of a second order $\mathbb{Z}_2$-breaking transition. The number of bubbles per hubble patch at temperature $T$ in general reads
\begin{align}
	\mathcal{N}_{\, \rm DW} (T) = \int_{T}^{T_c} \frac{ dT'}{T'}
	 \, \frac{ \xi  \Gamma_{\rm DW}(T')}{H(T')^{3}} \, ,
	\label{eq:nucleationseed}
\end{align}
where $\xi H^{-2}(T)$ is the average domain wall surface area in one Hubble patch. As in the homogeneous case, we can define a nucleation temperature for the catalysed transition via the condition $\mathcal{N}_{\, \rm DW} (T_\nuc) =1$. 
This results in $B_\DW \simeq 105$ at the electroweak scale for $\xi = 1$ (with a weak logarithmic dependence on $\xi$). 
\medskip

Let us finally emphasise that the nucleation temperature $T_\nuc$ defined through equations~\eqref{eq:nucleationhom}~and~\eqref{eq:nucleationseed} 
indicate the temperature at which there is expected to be one bubble in each hubble volume. By itself this condition being fulfilled does not guarantee that the phase transition will actually complete in the expanding universe. This discussion on percolation is postponed to section~\ref{subsec:percolation}.

\section{Numerical analysis of the catalysed transition}
\label{sec:MPT}

\subsection{The mountain pass algorithm}

\label{sec:MPTdetail}

Solving the system in \eqref{eq:eomgen} directly is challenging as the
spherical symmetry in equation~\eqref{eq:hombubble} is broken by the
presence of the domain wall. In ref.\,\cite{Blasi:2022woz}, this issue
was circumvented by employing a Kaluza--Klein (KK) decomposition along the
$z$ direction and reducing the 4d theory to a 3d EFT
written in terms of the KK modes. This reduces the problem of seeded tunnelling to that of
homogeneous tunnelling within a lower dimensional theory living on the
domain wall, which can be solved with standard techniques. However, this method becomes technically more difficult to apply beyond the high--$T$ approximation as there is no analytical form for the domain wall profile or the KK modes. In addition, the effective theory on the domain wall may not be under control in the whole temperature range relevant for the tunnelling. 

\medskip

In order to determine the phase transition dynamics beyond the high--$T$ expansion, and to study all the relevant temperature range, we will here look for solutions of \eqref{eq:eomgen} by following the strategy of ref.\,\cite{Agrawal:2022hnf}. The algorithm presented in~\cite{Agrawal:2022hnf} is based on the mountain pass theorem~\cite{AMBROSETTI1973349}, which turns the issue of solving the PDEs into a minimization problem that can be more easily tackled numerically. The algorithm is initialized by constructing a path in field space that connects the false vacuum configuration to a supercritical bubble profile with a smaller euclidean action. In this case, the false vacuum is
\be
	S_\fv(\rho, z) = S_\text{DW}(z), \quad h_\fv(\rho, z) \equiv 0 \, .
\ee
and we use the following ansatz for the supercritical bubble: 
\begin{align}
	S_{\SC} (\rho, z) &=\frac12 \left[  S_{\rm DW} (z+ f(\rho/R_0) Z_0)  +  S_{\rm DW} (z - f(\rho/R_0) Z_0) \right] \, ,
	\\
	\quad h_{\SC} (\rho, z) &= \frac{v_h}{2}\left[ 1 - \tanh \left( \frac{(\rho/R_0)^2 + (z/Z_0)^2-1}{\delta}\right) \right] \, ,
\end{align}
where $R_0$ sets the bubble radius in the $r$ direction, $Z_0$ sets the radius in the $z$ direction and $\delta$ sets the thickness of the higgs profile. The function $f(x)$ we take to be equal to
\be
	f(x) = 1- \tanh(x^4)
\ee
for parameters in the thin wall regime (tunnelling close to the critical temperature), or $f(x) = e^{-x^2}$ at temperatures significantly below the critical temperature. In each case we vary the parameters $R_0, Z_0, \delta$ to find a profile which gives 
an action smaller than the one in the false vacuum.

\medskip

Here we describe the basic features of the mountain pass algorithm, but refer the reader to ref.~\cite{Agrawal:2022hnf} for more details. After determining the false vacuum and supercritical profiles we define a path in field space between the points $(h_\fv, S_\fv)$ and $(h_\SC, S_\SC)$. We find the profile along the path which maximises the action, $(h_{\rm max}, S_{\rm max})$, then update the profiles using a gradient descent algorithm to minimise a cost function. The cost function we choose is a measure of how well the equations of motion are satisfied by the profiles $(h_{\rm max}, S_{\rm max})$:
\begin{align}
	C &= \sum_{ij} \left( E[h_{\rm max} (\rho_i, z_j)]^2 +  E[S_{\rm max}  (\rho_i, z_j)]^2 
	\right) \, ,
	\\
	E[\phi(\rho, z)] &=\frac{\partial^2 \phi}{\partial \rho^2} + \frac{1}{\rho} \frac{\partial \phi}{\partial \rho} + \frac{\partial^2 \phi}{\partial z^2} - \frac{\partial}{\partial \phi} V_\text{eff}(h,S;T)
	\, ,
\end{align}
where the sum is over a grid of points $(\rho_i, z_j)$. We note that this differs from the implementation of the algorithm in~\cite{Agrawal:2022hnf} where the cost function used was the action itself. 

\begin{figure}
\begin{tabular}{lr}
	\includegraphics[scale=0.4]{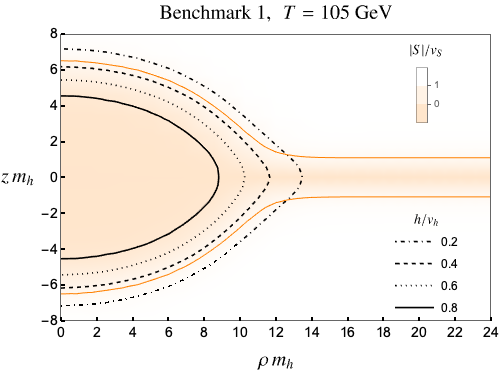}
	&
	\includegraphics[scale=0.4]{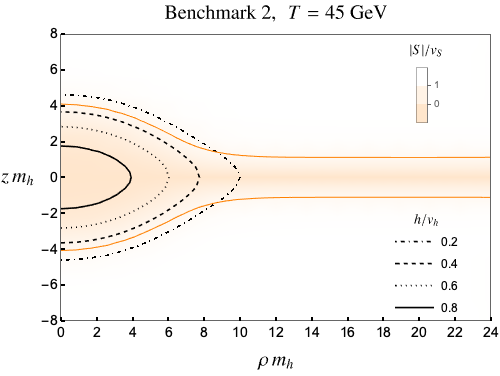}
\end{tabular}

\caption{Field profiles showing the critical bubble nucleated on the domain wall for benchmark 1 (left) and benchmark 2 (right). The shading represents the value $|S|/v_S$ and black contours show lines of constant $h$. The orange lines show the contours $|S|/v_S= 0.75$.}
\label{fig:profiles}
\end{figure}

The cost function $C$ is evidently minimised when $(h_{\rm max}, S_{\rm max})$ solve the equations of motion, but there are many solutions: the domain wall profile, the constant solutions $(h, S) = (v_h, 0)$ and $(h, S) = (0, v_s)$, the homogeneous tunnelling solution and the domain wall bounce solution we are interested in. The use of the mountain pass algorithm ensures that we find the domain wall bounce solution by restricting the field configurations considered to lie on the path between $(h_\fv, S_\fv)$ and $(h_\SC, S_\SC)$.
Examples of the field profiles for critical bubbles found using the mountain pass algorithm are shown in figure~\ref{fig:profiles} and display the expected elliptical shape.

\subsection{Results for the bounce action }
\label{sec:results}

With the numerical procedures described in the previous section we are now in a position to compare the bounce solutions for the catalysed and homogeneous configurations. If we restrict ourselves to the leading order in the high--$T$ expansion we can crosscheck the results previously presented in ref.\,\cite{Blasi:2022woz} with the new ones obtained with the MPT. This is shown in appendix~\ref{sec:highT} for a given benchmark point, and we find a very good agreement between the various methods in their region of applicability.

In this section we instead focus on the results obtained with the MPT with the full 1--loop thermal potential. To this end, we consider two benchmark points in parameter space, indicated by the red and blue stars in figure~\ref{fig:scan}. 
Benchmark 1 is the following choice of parameters (corresponding to benchmark C in section 4.2.2 of \cite{Caprini:2015zlo}): 
\be
(\kappa,\eta, m_S) = (1.5,\, 3.3,\, 250 \, \GeV)\, ,
\ee
and benchmark 2 is the choice:
\be
(\kappa, \eta, m_S) = (1.5, \, 2.3, \, 250 \, \GeV)\, .
\ee
In figure~\ref{fig:profiles} we show the critical bubbles at nucleation for both benchmarks.

\begin{figure}
\begin{tabular}{lr}
\includegraphics[scale=0.27]{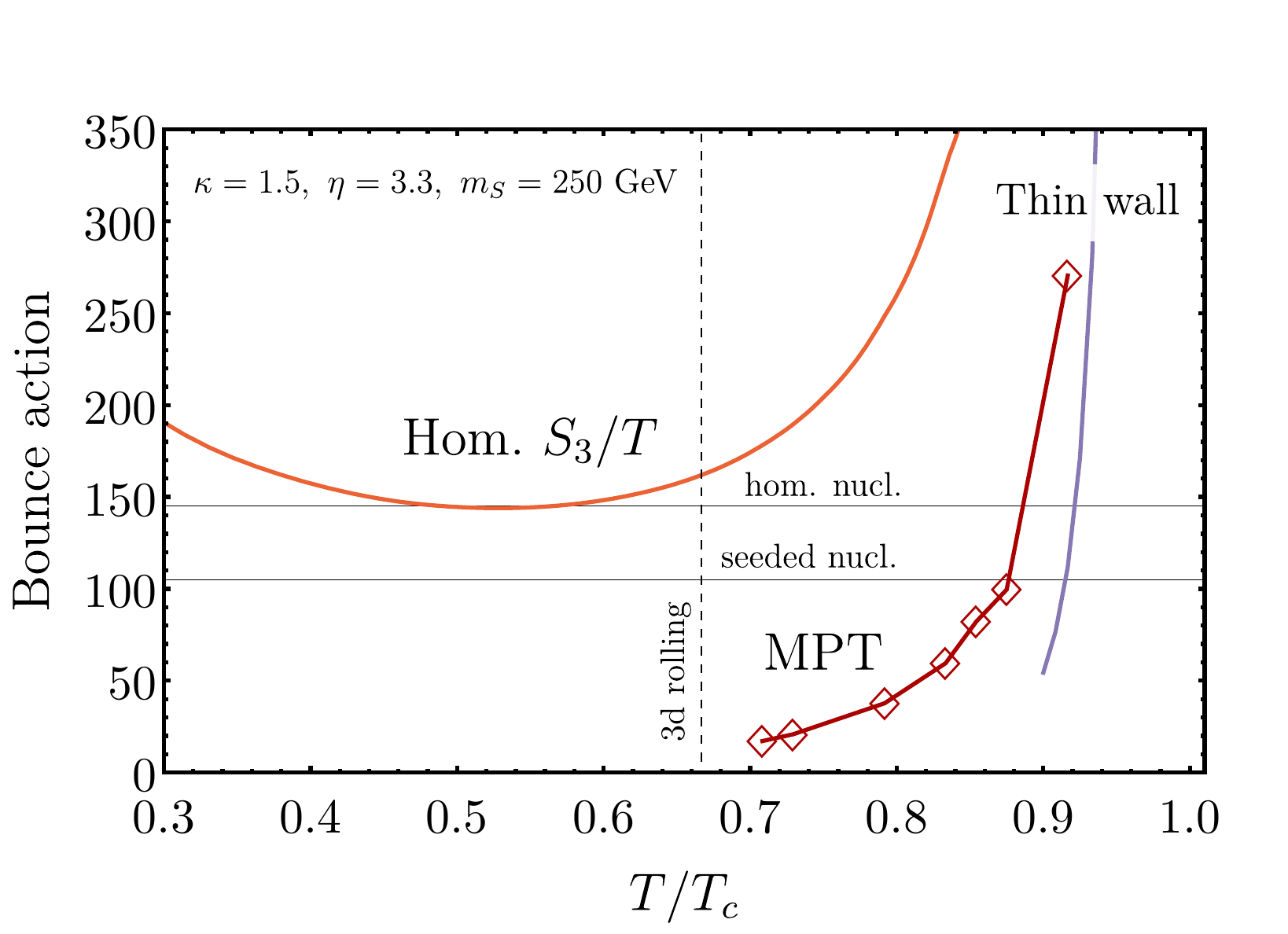}
&
\includegraphics[scale=0.273]{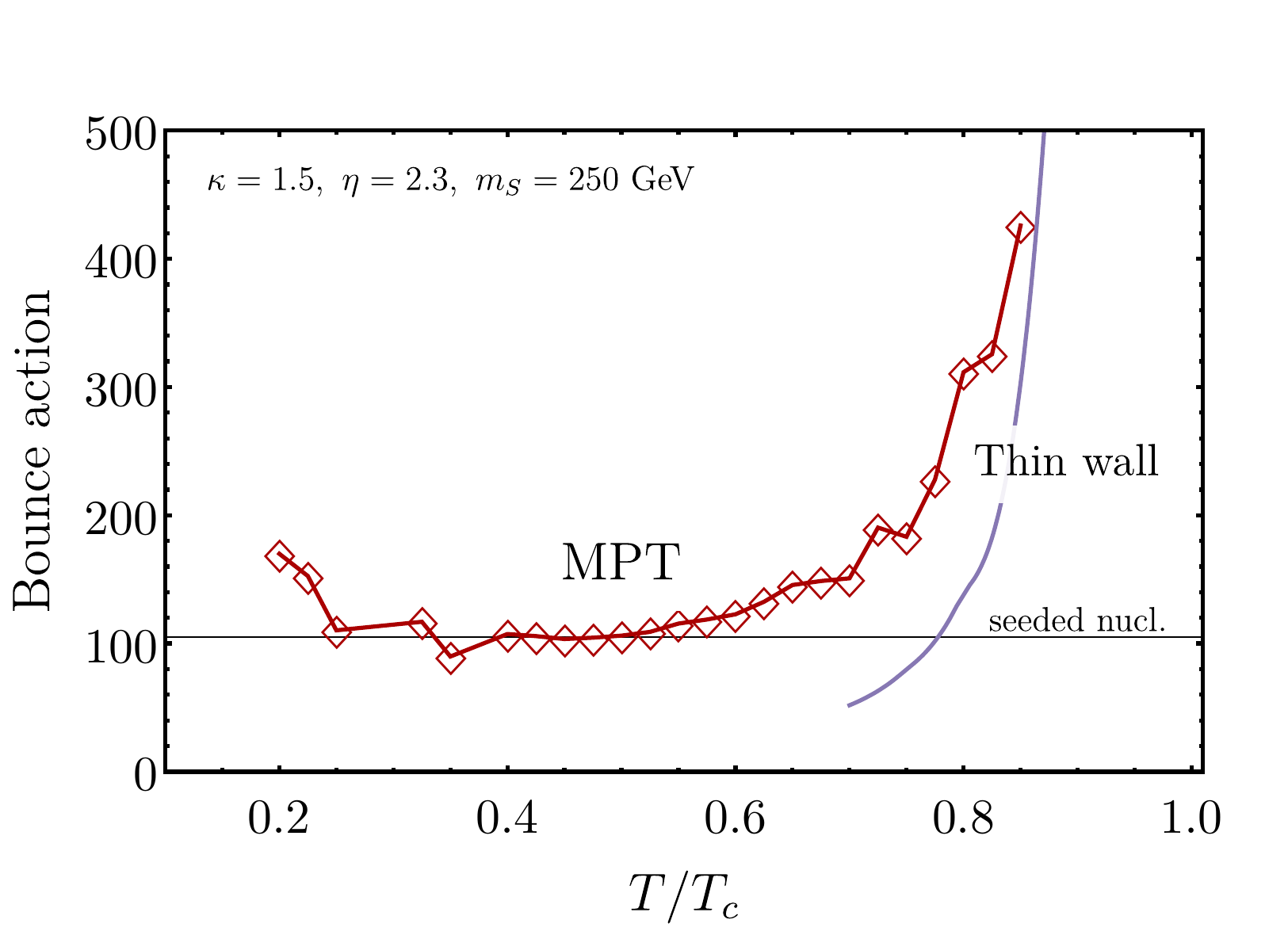}
\end{tabular}
\caption{Comparison of the bounce actions for the homogeneous and catalysed transitions as a function of temperature for both of the benchmark points. For benchmark 2 (right panel) the homogeneous bounce action is not shown because the values are $B_{\hom}\sim 800$, far above the value required for the phase transition to complete. The purple lines show a lower bound on the seeded bounce action evaluated within the thin wall approximation, see appendix~\ref{app:ThinWall}. }
\label{fig:bench} 
\end{figure}

In figure~\ref{fig:bench} the bounce action for the catalysed transition is compared to the homogeneous bounce action for a range of temperatures for each benchmark. Also shown is a lower bound on the bounce action derived within the thin wall approximation, and therefore reliable for $T \approx T_c$, as detailed in appendix~\ref{app:ThinWall}. For both benchmark points, the catalysed bounce action is much smaller than the homogeneous bounce, $B_\DW < B_{\hom}$. 
Given the exponential dependence of the decay rate on the bounce action, at a given temperature the catalysed process will dominate the homogeneous decay,
\begin{align}
	\Gamma_{\hom} (T) \ll \xi H(T) \Gamma_\DW (T) \, .
\end{align}
As the catalysed rate depends only linearly on the number of domain walls per hubble patch, $\xi$, this is the case even for a sparse domain wall network ($\xi \sim 1$). 

For benchmark 1 the bounce action actually vanishes at low temperatures, meaning that the barrier for the seeded tunneling is thermal and disappears around $T/T_c \simeq 0.67$ (as indicated by the dashed vertical line by ``3d rolling"). This temperature can be found by referring to the mass of the lightest Higgs mode living on the wall, $\omega_0^2$\,\cite{Blasi:2022woz}.
When thermal corrections are included beyond the high--$T$ expansion, $\omega_0^2$ can only be obtained numerically by solving the eigenvalue equation
\be
-\phi_0^{\prime\prime}(z) + \frac{\partial^2 V_{\rm eff}(h, S_{\rm DW}(z);T)}{\partial h^2}\bigg|_{h=0} \phi_0(z) = \omega_0^2(T) \phi_0(z).
\label{eq:omega_0}
\ee
The temperature for which the barrier disappears is then found by setting $\omega_0^2(T)=0$ so that the domain wall is classically unstable.

Also shown in figure~\ref{fig:bench} are the nucleation conditions for the seeded and homogeneous transitions (assuming radiation domination) identified in section~\ref{sec:setup}. In both cases, this condition identifies the temperature at which there is on average one bubble of true vacuum per Hubble patch. As we can see, homogeneous nucleation barely occurs in benchmark 1, and similarly for seeded nucleation in benchmark 2. These two benchmark points are also shown in figure~\ref{fig:scan} and sit at the boundaries of the red and blue regions, respectively.
\\

The two benchmark points represent examples of the different effects domain wall catalysis can have on the phase transition. Benchmark 1 is a point where the homogeneous transition only just completes ($\Gamma_{\hom} (T) =H^4(T)$ at the minimum of $B(T)$) but the catalysed transition happens much earlier. The presence of domain walls means $T_\nuc$ is much closer to the critical temperature than in the homogeneous case. Benchmark 2 is an example case where the homogeneous transition is too slow to complete (equation~\eqref{eq:nucleationhom} is not satisfied) but the catalysed process does finish. This point in parameter space would be naively ruled out by the strongly suppressed 
homogeneous phase transition, but is viable if domain walls are present.

\section{Phenomenology of the phase transition}
\label{sec:PTparams}

In this section we discuss the different phenomenological properties of homogeneous and seeded phase transitions in terms of the percolation conditions, size of bubbles and energy released in the transition. These properties are related to (but also go beyond) the difference in the nucleation temperature discussed in the previous section, and can have important implications for the observational prospects in terms of gravitational waves, as we shall discuss in section~\ref{sec:gravwaves}.

\subsection{Time scale}
In the case where nucleation does occur for a homogeneous phase transition, we can define a characteristic timescale, $\beta^{-1}$, given by:
\begin{align}
	\beta &= - \left . \frac{d B_{\hom}}{dt}\right |_{T_\nuc}  
	=H_\nuc T_\nuc
	\left . \frac{d B_{\hom}}{dT}\right |_{T_\nuc} \, ,
\end{align}
where $H_\nuc = H(T_\nuc)$. This quantity sets the average time for two bubbles to collide after they are nucleated. 

For a catalysed transition, a similar timescale can be calculated with $B_\DW$ in place of $B_{\hom}$. However, for sparse domain wall networks ($\xi \sim 1$) the time taken for two bubbles to collide will instead be set
by the average distance between the walls as $(v_w \xi H_\nuc)^{-1}$, where $v_w$ is the bubble wall velocity. The effective inverse time scale for the catalysed phase transition will then be given by the slowest (bottleneck) process, namely
\begin{align}
	\beta_\DW =
	\min \left(\left . \frac{d B_{\DW}}{dt}\right |_{T_\nuc} , 
	\, v_w \xi H_\nuc \right) \, .
	\label{eq:timescales}
\end{align}

\subsection{Percolation temperature}
\label{subsec:percolation}

It is useful to define a percolation temperature, $T_\perc$, that captures the moment at which a significant fraction of space is filled with the true vacuum and the transition is about to complete. 
To this end, it is customary to refer to the probability $p(T)$ of finding a point that still remains in the false vacuum at temperatures $T$ below the critical temperature $T_c$~\cite{Guth:1981uk,Guth:1982pn}:
\be 
\begin{aligned}
\label{eq:3Dperc}
 	p (T) &= e^{-I(T)}, \quad 
	I (T) &= \frac{4 \pi}{3} \int^{T_c}_{T}  dT' \,  
	\frac{ \Gamma(T^\prime) a(T^\prime)^3 r(T,T^\prime)^3 }{H (T') T'} \,,
\end{aligned}
\ee
where $\Gamma(T)$ is a homogeneous nucleation rate per unit volume,
and $r(T,T^\prime)$ is the (co-moving) size at the time $t(T)$ of a spherical bubble nucleated at the time $t^\prime(T^\prime)$,
\be
	\label{eq:radius}
	r(T, T^\prime) = \int^{T^\prime}_T \frac{d \tilde T}{\tilde T} \, \frac{v_w}{  H(\tilde T)  a(\tilde T) } \, .
\ee
The quantity $I(T)$ corresponds to the total volume of spherical true vacuum bubbles (overcounting overlapping regions) per unit comoving volume at the temperature $T$.

For the case of spherical bubbles in 3D space, percolation is assumed to occur when $I(T_\perc) = n_c \simeq 0.34$ and $p(T_\perc) \simeq 0.71$~\cite{Guth:1981uk,Guth:1982pn, PhysRevD.46.2384}. For a homogeneous phase transition the percolation temperature can then be evaluated straightforwardly according to \eqref{eq:3Dperc} by taking $\Gamma= \Gamma_{\rm hom}$ in \eqref{eq:homrate}.

For the catalyzed process the percolation temperature depends on which of the two possible time scales in Eq.\,\eqref{eq:timescales} controls the duration of the transition.
For a very dense network with $\xi \gg 1$, or a scenario in which the (seeded) bounce action is very flat around the nucleation temperature, we expect the time scale of the transition to be set by the slope of the bounce action. This makes the percolation condition for this scenario equivalent to a homogeneous phase transition, and the percolation temperature may then be computed according to \eqref{eq:3Dperc} with the following identification:
\be
	\label{eq:gammaxi}
	\Gamma  = \xi H  \Gamma_{\rm DW} ,
\ee
where $\Gamma_{\rm DW}$ is given in \eqref{eq:DWrate}. We shall indicate by $T_\perc^{\rm 3D}$ the percolation temperature obtained in this way.

On the other hand, whenever the distance among domain walls sets the relevant time scale (namely for sparse networks with $\xi = \mathcal{O}(1)$ or for very fast nucleation rate) the percolation temperature can be derived from a purely geometrical argument, similarly to what has been done for a distribution of monopoles\,\cite{Guth:1981uk}. In this case bubbles of true vacuum can be thought of growing instantaneously out of the domain walls at $T \simeq T_\nuc$. From that moment on, the amount of true vacuum volume per unit co-moving volume is given by
\be
\begin{aligned}
	\label{eq:percseed}
	I_{\rm w}(T) =
       (\xi H a)_\nuc \cdot 2 r(T, T_\nuc),
\end{aligned}
\ee
where $(a H)_\nuc^{-1}$ is the co-moving Hubble radius at $T=T_\nuc$. The probability of a point remaining in the false vacuum is $p_{\rm w}(T) = e^{-I_{\rm w}(T)}$, so that $I_{\rm w}(T)$ is the equivalent of $I(T)$ for a domain--wall seeded transition. The derivation of \eqref{eq:percseed} is presented in appendix~\ref{sec:percgen}, together with the analogous derivation for a network of monopoles and strings acting as nucleation sites.
The percolation temperature for this case, $T_\perc^\xi$, is then defined through $I_{\rm w}(T_\perc^\xi) = n_c$\,\footnote{In the following, we shall keep $n_c = 0.34$ also for the case of sparse networks, even though this value refers to spherical bubbles with a homogeneous nucleation probability.}.
Assuming radiation domination, one finds
\be 
	\label{eq:geomperc}
	T_\perc^\xi = \frac{1}{1 + \frac{n_c}{2 \xi v_w}} T_\nuc,
\ee
which for relativistic wall velocities and $\xi \sim 1$ gives $T_\perc \sim 0.85\, T_\nuc $. In general, we define the percolation temperature for the catalyzed transition as
\be
	T_\perc = \,\text{min}(T_\perc^{\rm 3D}, T_\perc^\xi).
\ee

\medskip

In the case of strong supercooling one needs to impose an additional condition on top of percolation to make sure that the phase transition does actually complete. This condition requires the \emph{physical} volume that still remains in the false vacuum to decrease with time~\cite{PhysRevD.46.2384}. This requirement applies to both homogeneous and seeded phase transitions, and becomes particularly relevant when percolation occurs close to the moment where the vacuum energy of the false vacuum starts to dominate the energy budget of the universe. 
The condition that the false vacuum volume decreases reads
\be
	\label{eq:volumedecrease0}
	3 H -  \frac{d I }{d t} < 0 \, ,
\ee 
where the appropriate $I(T)$ depends on the nature of the transition as discussed above. This condition is usually checked at $T = T_\perc$, see  e.g.\,\cite{Ellis:2018mja}.

For instance, in the case of a sparse network where $I = I_{\rm w}$ in \eqref{eq:percseed},
by evaluating Eq.\,\eqref{eq:volumedecrease0} at percolation, we obtain
\begin{align}
	\label{eq:volumedecrease}
	3 - 2 v_w \xi \,\frac{H(T_\nuc)}{H(T_\perc)} \frac{T_\perc}{T_\nuc} & < 0.
\end{align}
If we further assume that percolation is happening sufficiently far from vacuum domination we can use \eqref{eq:geomperc} for $T_\perc$, and the condition above simplifies to
\be
2 v_w \xi + n_c > 3.
\ee
As we can see, for $\xi \leq (3-n_c)/2v_w \simeq 1.33$, the volume in the false vacuum may not decrease at $T_\perc$, but only at some lower temperature (provided that vacuum domination has not set in yet). 
In the following, we shall consider $\xi \geq 2$ so that \eqref{eq:volumedecrease} is satisfied at $T_\perc$ provided that $v_w \sim 1$. 

\subsection{Bubble sizes}
\label{sec:bubblesize}

The timescale of the phase transition sets the average size of
bubbles at percolation, $R_\perc$. For homogeneous transition we shall
refer to the standard relation
\be
\label{eq:Rast}
R_\perc^{\, \rm hom} = - v_w (8\pi)^{1/3} \left( \left . \frac{d B_{\hom}}{dt}\right |_{T_\perc} \right)^{-1}.
\ee
The properties of a seeded phase transition whose time scale is set by the slope of
the bounce action are indistinguishable from a homogeneous phase
transition. The average bubble size is then still given by
\eqref{eq:Rast} with $B_{\hom}$ replaced by $B_{\rm DW}$:
\be 
\label{eq:Rast1}
R_\perc^{\, \rm 3D} = - v_w (8\pi)^{1/3} \left( \left . \frac{d B_{\rm DW}}{dt}\right |_{T_\perc} \right)^{-1}.
\ee
For sparse networks, however, the bubble size will be related to the average distance among the walls at the nucleation temperature. We then find
\be
\label{eq:Rast2}
R_\perc^{\, \xi} = a(T_\perc) r(T_\perc, T_\nuc) = \frac{n_c}{2\xi H(T_\nuc)}
\left( 1 + \frac{n_c}{2 \xi v_w} \right),
\ee
where in the last step we have used the percolation condition
$I(T_\perc^\xi)=n_c$ together with the relation \eqref{eq:geomperc}, which assumes percolation during radiation domination. The last equality may be modified close to vacuum domination.

Considering then both the possibility of sparse and dense networks,
the appropriate value of the bubble size for the catalyzed phase transition will be taken to be the largest of \eqref{eq:Rast1} and \eqref{eq:Rast2},
\be
\label{eq:Rastseed}
R_\perc^{\, \rm seed} = \,{\rm Max} ( R_\perc^{\, \rm 3D}, R_\perc^{\, \xi} ).
\ee

\medskip
\subsection{Latent heat}
\label{sec:heat}

Another important quantity that characterises the transition is the energy released in the plasma, normalised to the radiation energy density at $T_\perc$. This quantity is usually denoted by $\alpha$ and is given by (see e.g.\,\cite{Ellis:2020nnr})
\be
	\alpha = 
	\frac{1}{\rho_\text{rad}(T_\perc)} 
	\left( \Delta V - \frac{T}{4} \frac{ \text{d} \Delta V}{\text{d} T} \right)\bigg |_{T = T_\perc}\, ,
	\label{eq:alpha}
\ee
where $\Delta V$ is the difference in energy between the true and false vacua at $T=T_\perc$. The relation in \eqref{eq:alpha} applies both to homogeneous and catalyzed phase transitions.

\subsection{A benchmark analysis}
\label{sec:benchmarkanalysis}

To make the discussion in the previous sections more concrete, we determine the quantities discussed above for a representative set of parameters: $\eta = 1$, $m_S = 200 \,{\rm GeV}$, with the portal coupling $\kappa$ free to vary, as considered also in ref.\,\cite{Ellis:2018mja}. For simplicity we shall take $v_w \sim 1$ throughout this section.

In order to efficiently scan different values of $\kappa$, we use the thin wall approximation (whose details are presented in appendix~\ref{app:ThinWall}) to evaluate the seeded bounce action. As shown in figure~\ref{fig:bench}, this approximation captures the overall qualitative behaviour of the bounce action compared to the fully--numerical result. The thin wall approximation gives a good approximation to the nucleation temperature when the transition completes close to $T_c$, but becomes worse for more supercooled transitions. The thin wall approximation also predicts a larger slope for the bounce action at the nucleation temperature, however, for sparse domain wall networks this slope is irrelevant as the bubble size is set by the domain wall separation. We then expect our results to give a good indication of the difference between the catalysed and homogeneous transitions.

\begin{figure}
	\centering
	\includegraphics[scale=.27]{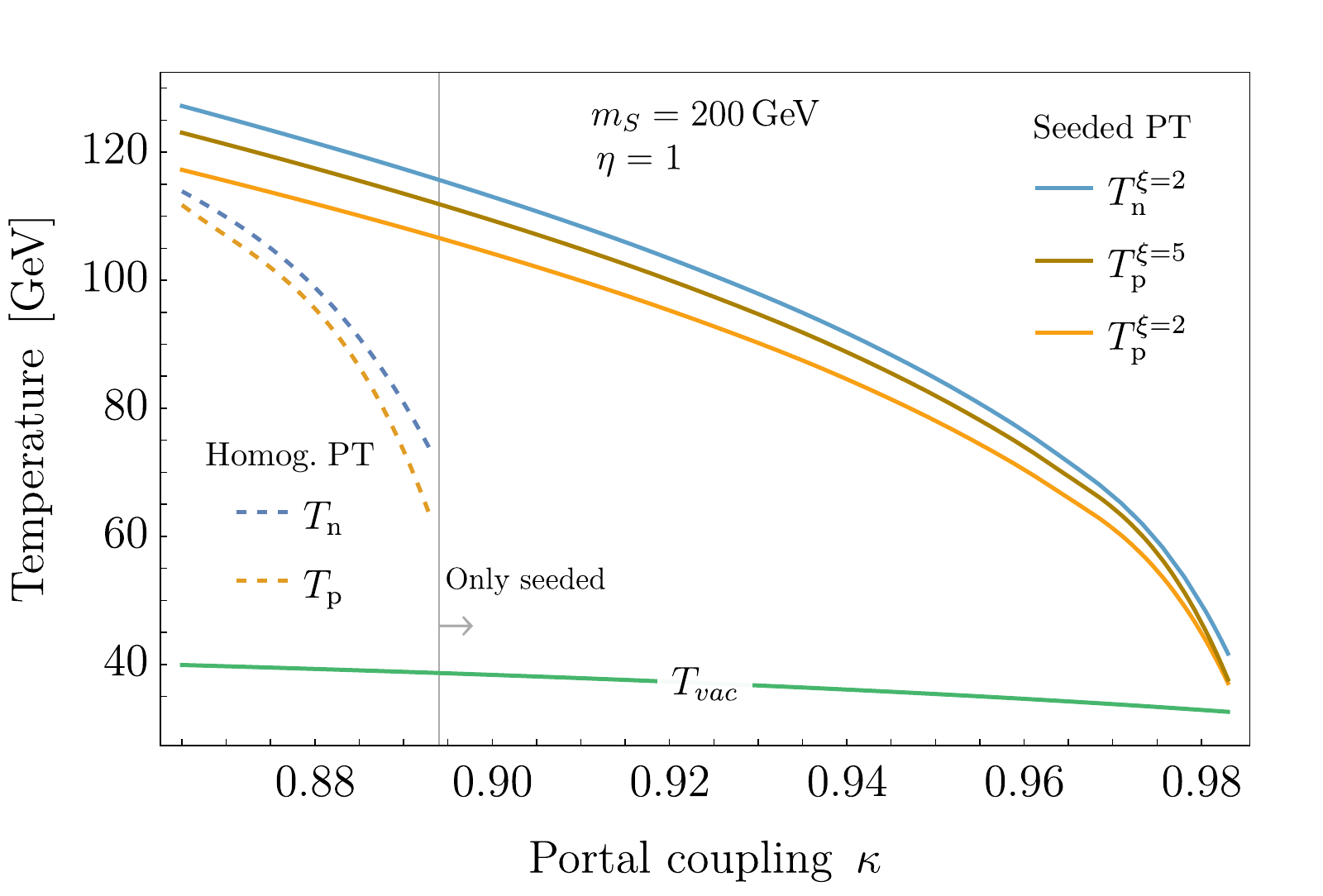}
	\includegraphics[scale=.27]{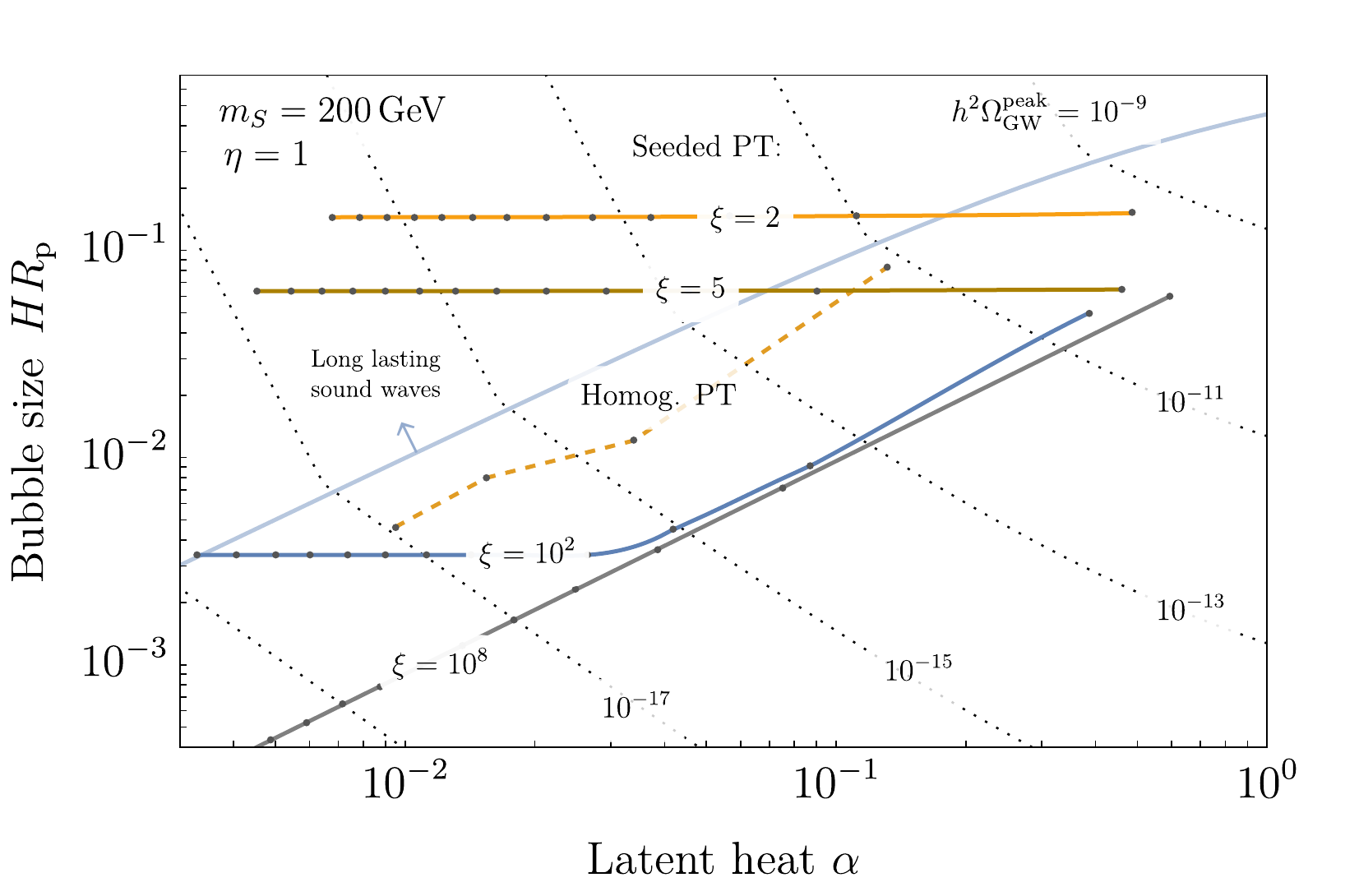}
	\caption{ \textbf{Left.} Nucleation and percolation temperature, $T_\nuc$ and $T_\perc$, for the homogeneous phase transition (dashed) and the seeded phase transition (solid) for different values of $\xi$ as a function of the portal coupling $\kappa$ fixing $\eta=1$ and $m_S= 200\, \rm GeV$. The green line indicates the temperature at which the vacuum energy begins to dominate.
\textbf{Right.} Trajectories in the $(\alpha, H R_\perc)$ plane by varying $\kappa$ according to  the range in the left panel ($\kappa$ increases moving from left to right) for the homogeneous transition (dashed) and the seeded transition (solid) for different values of $\xi$. Also shown are contours indicating the peak gravitational wave amplitude from sound waves, as well as the region where sound waves can be long lasting (see  section~\ref{sec:gravwaves} for more details). The wall velocity has been fixed to $v_w \sim 1$ in both panels.}
	\label{fig:homvsseed}
\end{figure}

The nucleation and percolation temperatures for the second step of the phase transition, $(\pm v_s,0) \rightarrow (0,v_h)$, as a function of $\kappa$ are shown in the left panel of figure~\ref{fig:homvsseed}, for homogeneous and seeded nucleation by the dashed and solid lines, respectively. For the latter, different numbers of domain walls per Hubble patch are considered, namely $\xi =2$ and $\xi = 5$.

As we can see, the homogeneous phase transition becomes more and more supercooled for larger $\kappa$, until it fails to complete for $\kappa \gtrsim 0.90$. On the other hand, we find successful percolation for the seeded phase transition beyond this value of the portal up to $\kappa \approx 0.98$, in agreement with the general behaviour discussed around figure~\ref{fig:scan}. As we are using the thin wall approximation in this analysis, this value of the end point should however be taken only as an indication.

Due to the weak logarithmic dependence on $\xi$, the nucleation temperature defined in equation~\eqref{eq:nucleationseed} is practically the same for all the considered values of $\xi$, and we then only show the case of $\xi=2$. On the other hand, the percolation temperature $T_\perc$ has a relatively stronger dependence on $\xi$ as long as the network is sparse, $\xi \sim \mathcal{O}(1)$, as shown explicitly in figure~\ref{fig:homvsseed} left. While the dependence of $T_\perc$ on $\xi$ is still mild, this can lead to a significant change in other quantities such as $\alpha$ and $R_\perc$, as shown by the different trajectories in the right panel of figure~\ref{fig:homvsseed}. For larger values of $\xi$, $T_\perc$ and $T_\nuc$ are practically the same (as long as the relevant time scale of the transition is set by the domain wall distance).

The bubble size, $H R_\perc$, and latent heat, $\alpha$, as $\kappa$ is varied are shown in the right panel of figure~\ref{fig:homvsseed} for both the homogeneous and the seeded transition (by the dashed and solid trajectories, respectively).
The black dots indicate the sampled values of the portal coupling: they are uniformly sampled with $\Delta \kappa = 0.01$ within $\kappa \in (0.865,0.895)$ for the homogeneous transition, and within $\kappa \in (0.865,0.985)$ for the seeded counterpart. The values of $\alpha$ and $R_\perc$ are evaluated according to  section~\ref{sec:bubblesize} and  section~\ref{sec:heat}.

As we can see, the trajectory for the homogeneous phase transition shows a strong correlation between $\alpha$ and $R_\perc$, so that larger $\alpha$ implies larger bubbles, in agreement with previous studies (see e.g.\,\cite{Ellis:2018mja}).
In this respect, the seeded phase transition shows a very different behaviour, rooted in the fact that the relevant time scale may be set by the distance among the walls rather than the slope of the bounce action. This is in fact the case for $\xi \lesssim \mathcal{O}(10)$ for all the relevant values of $\kappa$, and the bubble size is given by $H R_\perc  \propto 1/\xi$ independently of $\alpha$. This simple scaling with $\xi$ is apparent for the trajectories with $\xi=2$ and $\xi = 5$, and also for the flat section of the one with $\xi = 10^2$.

For larger values of $\xi$, the slope of the bounce action may set the time scale, so that $R_\perc^{\,\rm 3D} > R_\perc^{\, \xi}$ in Eq.\,\eqref{eq:Rastseed}. In this case the correlation between $R_\perc$ and $\alpha$ typical of the homogeneous phase transition is naturally recovered. For $\xi = 10^2$, this happens for $\kappa \gtrsim 0.95$ and determines the sharp turn in the right panel of figure~\ref{fig:homvsseed}. For the extreme case of $\xi = 10^8$, corresponding to the densest possible domain wall network at the electroweak scale\,\footnote{This comes from a friction--dominated evolution of the network with $\sigma_{\rm DW}^{1/3} \sim T \sim 100 \,{\rm GeV}$, yielding $\xi \sim (100 \,\,{\rm GeV}/M_{\rm Pl})^{1/2}$~\cite{Vilenkin:2000jqa}.}, the time scale is always set by the slope of the bounce action for all the relevant values of $\kappa$. The phenomenology of the seeded phase transition becomes then practically independent of $\xi$, and indeed the trajectories for $\xi = 10^2$ and $\xi = 10^8$ approach each other in this regime. 
The implications of this behaviour for the gravitational wave signal expected from a seeded phase transition will be discussed in detail in the next section. 

\section{Gravitational wave signatures}
\label{sec:gravwaves}

In this section we discuss the gravitational wave signal from a seeded electroweak phase transition, and outline the corresponding detection prospects at LISA.

\subsection{Sound waves}
\label{subsec:soundwaves}

The key parameters determining the gravitational wave signal from a first order phase transition are its timescale, the latent heat, $\alpha$, and the bubble wall velocity, $v_w$. 
The first two quantities have been discussed in section~\ref{sec:PTparams}. As for the wall velocity, this is determined by balancing the vacuum energy driving the expansion of the bubble with the friction coming from the thermal plasma~\cite{Arnold:1993wc,Moore:1995ua, Moore:1995si, Bodeker:2009qy, Bodeker:2017cim}. 

For the case of the electroweak phase transition, $v_w$ will reach a terminal value due to transition radiation emitted as the massive gauge bosons cross the bubble wall which grows proportionally to the relativistic $\gamma(v_w)$ factor of the wall~\cite{Bodeker:2017cim}, see also refs.\,\cite{Gouttenoire:2021kjv,Azatov:2023xem} for recent studies. 
The main contribution to the gravitational waves comes then from sound waves in the plasma. The gravitational wave spectrum in terms of the average bubble size, $H_\perc R_*$, and the latent heat $\alpha$ is given by\,\cite{Hindmarsh:2013xza, Giblin:2014qia, Hindmarsh:2015qta, Hindmarsh:2016lnk, Hindmarsh:2017gnf, Cutting:2018tjt,Hindmarsh:2019phv, Lewicki:2021pgr}:
\begin{equation}
\begin{aligned}
\label{eq:GWsound}
	\frac{d \Omega_{\text{gw}\,,0}}{d \,\text{ln}(f)} &= 
	4.13 \times 10^{-7} \left( H_\perc R_*\right) 
	\left(\frac{\kappa_{\rm sw} \alpha}{1+\alpha}\right)^2 \left(\frac{100}{g_*}\right)^{1/3} S(f/f_{p,0}) \, ,
	\\
	f_{p,0} &\simeq 26 \left(\frac{1}{H_\perc R_*}\right)
	\left(\frac{T_\perc}{100 \, \GeV}\right)
	\left(\frac{g_*}{100}\right)^{1/6}
	\, \mu \text{Hz} \,,
	\\
	S(x) &= x^3 \left( \frac{7}{4+3x^2} \right)^{7/2},
\end{aligned}
\end{equation}
where $f_{p,0}$ is the peak frequency today.
The factor $K = \kappa_{\rm sw} \alpha/(1+\alpha)$ is the kinetic energy fraction of the fluid. It depends on the velocity profile of the fluid and is given by~\cite{Espinosa:2010hh}
\be
	K= \frac{4}{3} \bar{U}_f^2, \quad \bar{U}_f^2 = \frac{3}{v_w^3} \int_{c_s}^{v_w} \xi^2 \frac{v(\xi)^2}{1-v(\xi)^2} d\xi
	\, .
\ee
The velocity profile $v(\xi)$ is the solution of~\cite{Steinhardt:1981ct}
\be
	\frac{2 v}{\xi} = \frac{1 - \xi v}{1-v^2} \left[ \frac{1}{c_s^2} \left( \frac{\xi -v}{1-\xi v^2}\right)^2 -1 \right]
	\frac{d v}{d \xi},
\ee
where $c_s^2 = 1/3$ for a relativistic fluid,
with the boundary condition
\be
\label{eq:bcv}
	v(v_w) 
	= \frac{3\alpha}{2+3\alpha}.
\ee
Equation \eqref{eq:bcv} is only valid for relativistic wall velocities, $v_w \sim 1$, which is the case we will consider in what follows.

The gravitational wave spectrum in \eqref{eq:GWsound} is accurate only when the time for shock formation is larger than one Hubble time, namely if
\be
 	\frac{H_\perc R_*}{\bar U_f} = H_\perc R_* 
 	\sqrt{\frac{4( 1+\alpha)}{3 \kappa_{\rm sw}\alpha}} > 1.
 	\label{eq:SWcondition}
\ee
When this inequality is not satisfied, Eq.\,\eqref{eq:GWsound} needs to include an additional suppression factor given $H_\perc R_\ast/\bar U_f $\,\cite{Ellis:2018mja}\footnote{New contributions to the gravitational wave spectrum are then expected from the turbulent motion.}. The gravitational wave spectra presented above are strictly valid for rather fast phase transitions where the Hubble expansion can be neglected, whereas the signal from slower phase transitions may include additional suppression.

\subsection{Gravitational waves from seeded phase transitions}

Let us now apply the discussion of section~\ref{subsec:soundwaves} on the gravitational wave spectrum from a first order phase transition to the case of a catalysed process. First, we notice that domain wall catalysis is not expected to change the friction on the wall, so a difference in $v_w$ compared to the homogeneous case can only be due to a possible difference in $\alpha$. 

For the typical bubble size $R_\ast$ entering \eqref{eq:GWsound}, we use the size of bubbles at percolation, $R_\perc$, defined in equations~\eqref{eq:Rast} and~\eqref{eq:Rastseed} for the homogeneous and catalysed transitions respectively. For sparse networks, the connection between the correlation length of the domain wall network, $\xi_{\rm DW} = (\xi H)^{-1}$, and the effective size of the bubbles, Eq.\,\eqref{eq:Rast2}, has actually been observed in the numerical simulation of ref.\,\cite{Blasi:2023rqi}, where it was shown that the sound wave contribution to the gravitational wave spectrum can be obtained with good approximation by replacing the standard parameter $\beta$ with $1/\xi_{\rm DW}$. This justifies the use of $R_\ast = R_\perc^\xi \simeq (\xi H)^{-1}$ for sparse networks in our analysis \footnote{Notice that since the simulation is carried out assuming a negligible Hubble expansion, the application to domain wall networks with $\xi \lesssim 10$ requires extrapolation.}.

A novel aspect of seeded phase transitions is that the nucleated bubbles may deviate significantly from spherical symmetry, see figure~\ref{fig:profiles} and related discussion. This raises the question of what impact non--spherical bubbles could have on the gravitational wave signal. As the accelerated expansion of the bubble walls is driven by the vacuum energy released in the transition, the bubbles become more spherical as they expand~\cite{Adams:1989su, Garriga:1991ts, Garriga:1991tb}. As the gravitational wave signal is dominated by collisions between large (Hubble scale) bubbles which have expanded to many orders of magnitude larger than their initial size, the initial bubble profile has a negligible effect on the resulting gravitational wave signal.

However, the catalysed transition may lead to collisions of non-spherical bubbles as the nucleated bubbles are no longer uniformly distributed. This effect is most significant for sparse domain wall networks (small $\xi$), as bubbles are dominantly nucleated on domain walls which occupy only a small fraction of a hubble patch. In this case, bubbles will first merge to form a large, approximately planar, bubble before colliding with a bubble nucleated on a neighbouring domain wall. The gravitational wave spectrum from sound waves in this scenario was studied in ref.~\cite{Blasi:2023rqi}, where it was found that in fact the spectrum was unchanged from the homogeneous transition after the identification $R_\ast = (\xi H)^{-1}$.\footnote{A potential source of gravitational waves that this analysis neglects is a contribution from the acceleration of domain wall bubbles before they collide, as they are highly non-spherical. However, we do not consider this possibility in this work.}

In the following we shall only consider the standard sound wave contribution in Eq.\,\eqref{eq:GWsound} and capture the effect of the seeded transition via the use of the appropriate bubble size in Eq.\,\eqref{eq:Rastseed}.

\medskip

Some general implications of a seeded phase transition for the gravitational wave signal can already be inferred from the right panel of figure~\ref{fig:homvsseed}. To each point in the $(\alpha, H R_\ast)$ plane we can associate a gravitational wave peak amplitude according to Eq.\,\eqref{eq:GWsound}, and the various contours of constant $h^2\Omega_{\rm GW}^{\rm peak}$ are shown by the dotted black lines. As we can see, this amplitude strongly depends on the number of domain walls per Hubble volume, $\xi$, with larger $\xi$ implying a weaker signal.

\begin{figure}
	\begin{tabular}{lr}
		\includegraphics[scale=.5]{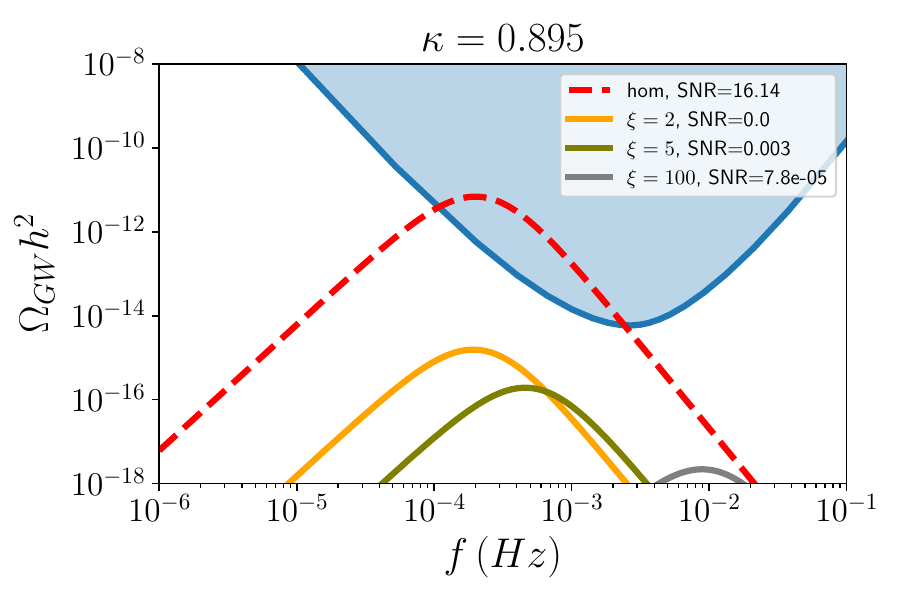}
		&
		\includegraphics[scale=.5]{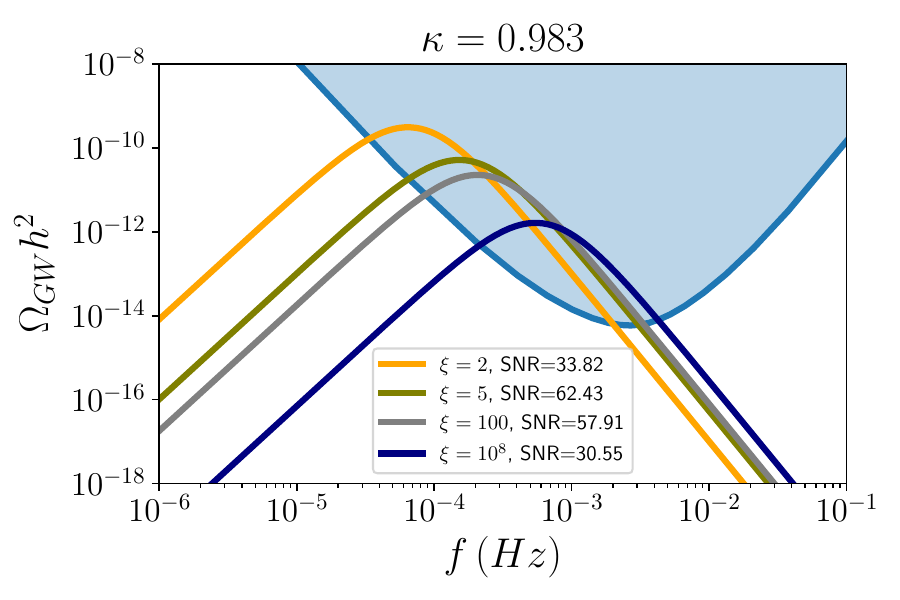}
	\end{tabular}
	\caption{Figures showing the gravitational wave signal for $\kappa = 0.895$ (left) and $\kappa =0.975$ (right). The orange curve, green, grey and blue curves show the catalysed transition for $\xi=1,5, 10^2, 10^8$ respectively. The red dashed curve shows the homogeneous transition, which only completes for $\kappa=0.895$. Blue shaded region shows the LISA sensitivity~\cite{Caprini:2019egz}.}
	\label{fig:GWsignal}
\end{figure}

Smaller values of $\xi$ result in larger bubbles and generically greater amplitudes. This, together with the decorrelation between $R_\perc$ and $\alpha$, allows the region of parameter space where sound waves are long lasting, i.e. where the inequality \eqref{eq:SWcondition} is satisfied, to be populated. The boundary of this region is indicated by the blue line in the right panel of figure~\ref{fig:homvsseed}. This condition is typically not met for homogeneous transitions because $H_\perc R_*$ and $\alpha$ both increase as the transition becomes supercooled~\cite{Ellis:2018mja}. This can also be seen explicitly from figure~\ref{fig:homvsseed} as the homogeneous trajectory for our benchmark always lies below the line of long--lasting sound waves.

A detailed comparison of the gravitational wave spectrum from each type of phase transition within the benchmark analysis of section~\ref{sec:benchmarkanalysis} is shown in figure~\ref{fig:GWsignal}. The power-law integrated sensitivity curves for LISA are shown in blue, computed using the procedure outlined in ref.~\cite{Thrane:2013oya} for a signal to noise ratio (SNR) of 1. The left panel shows the spectra for $\kappa=0.895$, which is the largest $\kappa$ value where the homogeneous transition still completes. In this case the expected signal from a homogeneous transition is larger than the catalysed transitions for all choices of $\xi$. This is because the catalysed transition percolates closer to the critical temperature, leading to less energy being released in the transition. The signal from a catalysed transition is larger for smaller $\xi$ values as the bubbles are larger at percolation. The $\xi =10^8$ transition is not shown as the signal is negligible.

The right panel of figure~\ref{fig:GWsignal} shows the signal for the catalysed transition at $\kappa = 0.983$, where the homogeneous transition does not complete but the seeded process is still fast enough. The signal for smaller $\xi$ values is larger and peaked at lower frequencies. For the $\xi=10^2, 10^8$ transitions the bubble size is no longer set by the domain wall separation but rather from the bubble nucleation rate, leading to a weaker signal peaked at higher frequencies. The gravitational wave signal is very sensitive to the precise value of $\kappa$ as we are close to the limit of successful percolation.

\begin{figure}
	\centering
	\includegraphics[scale=.35]{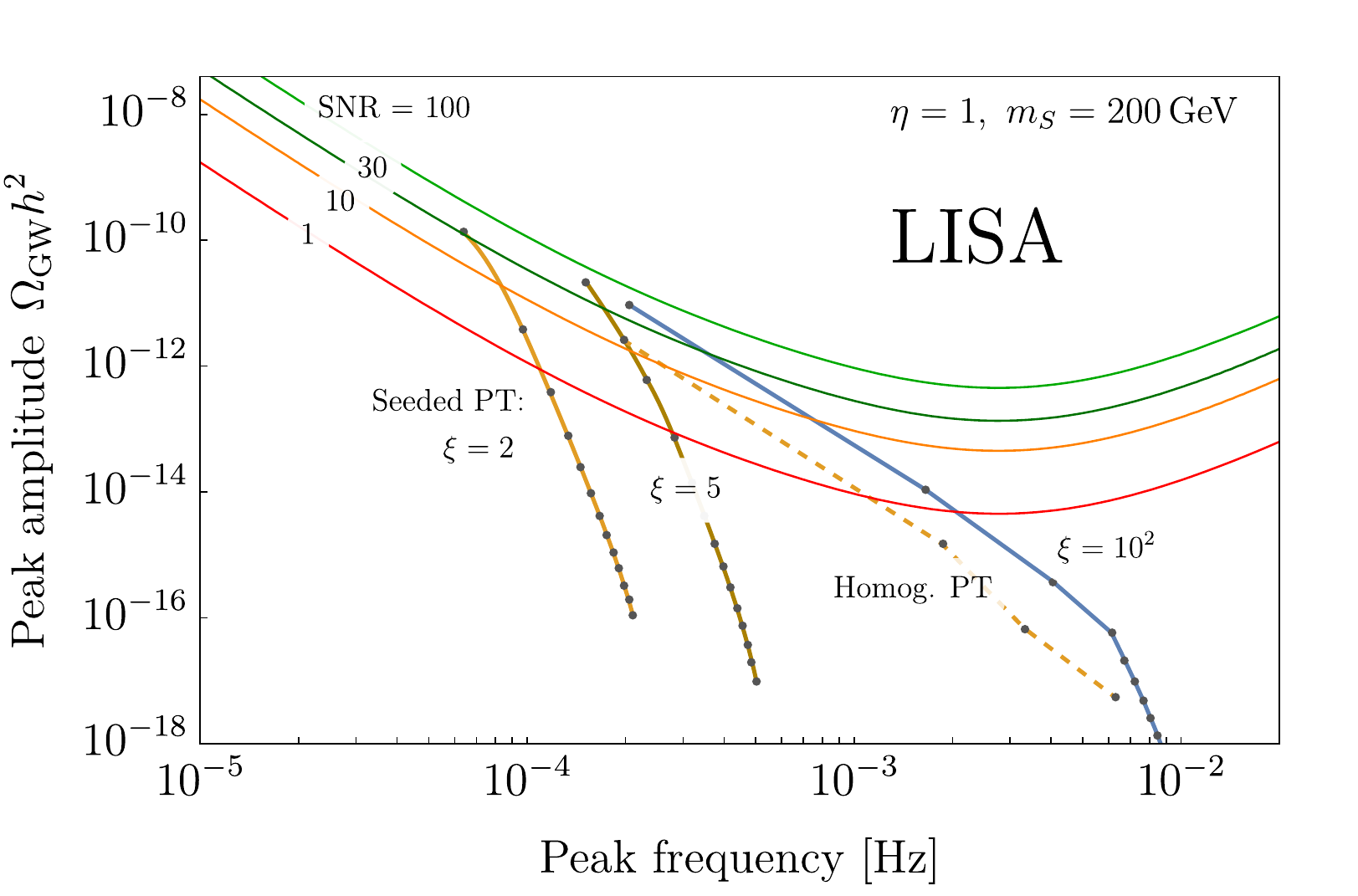}
	\caption{Peak frequency and amplitude of the gravitational wave signal from the homogeneous and catalysed transitions for the parameter points discussed in section~\ref{sec:benchmarkanalysis}. The solid lines indicate the catalysed transitions for different $\xi$ values and the dashed line shows the homogeneous transition. The coloured contours indicate the signal to noise ratio in LISA. }
	\label{fig:fpeakOmega}
\end{figure}

The gravitational wave spectrum as a function of $\kappa$ in the two--step parameter space for our representative benchmark of section~\ref{sec:benchmarkanalysis} can be summarised in figure~\ref{fig:fpeakOmega}, where we also highlight the implications for a detection at LISA. The different trajectories (solid and dashed for seeded and homogeneous phase transitions, respectively) are equivalent to the ones shown in figure~\ref{fig:homvsseed}, but are now plotted in terms of peak frequency and amplitude of the corresponding gravitational wave spectrum assuming again relativistic wall velocity. The black dots show the sampled values of the portal coupling, where $\kappa$ increases when moving upwards along the curves. 

The decorrelation of $\alpha$ and $R_\perc$ for seeded transitions with small $\xi$ manifests in figure~\ref{fig:fpeakOmega} as almost vertical trajectories. This is as $R_\perc$ is fixed by the domain wall separation, so the peak frequency shifts only due to the change in $T_\perc$ as $\kappa$ is varied, while the peak amplitude increases as $\Omega_{\rm gw} \propto \alpha^2 R_\perc$. This is in contrast to the homogeneous line and the transitions for $\xi = 10^2$ where the timescale is set instead by the nucleation rate and the peak frequency shifts more significantly due to the changing size of bubbles as they collide. Similarly to figure~\ref{fig:GWsignal}, a smaller $\xi$ generically leads to a larger signal amplitude. However, this does not necessarily translate in a larger SNR because of the shift to lower frequencies where LISA starts losing sensitivity. In this regard, future experiments such as $\mu$Ares\,\cite{Sesana:2019vho} could more efficiently probe the gravitational background in the case of sparse networks.

\section{Conclusions \& Outlook}
The xSM is a simple model which
encapsulates new physics that can modify the EWPT, possibly leading to an observable spectrum of GWs. The
$\mathbb{Z}_2$ symmetric limit of this model has been put forward as a
test case for simple weak scale new physics that can remain hidden from
current collider and precision searches, although it may be within reach of future colliders.

We have shown that in a large part of the relevant parameter space,
the phase transition dynamics are modified due to the presence of
domain walls. The decay of the false vacuum proceeds through
tunnelling catalyzed by domain walls, instead of through homogeneous
bubble nucleation. This changes the viability of specific parameter
space points in the model and qualitatively modifies the
gravitational wave signal.

The calculation of the bounce action and other parameters relevant for determining
gravitational wave spectra for catalyzed transitions are difficult
using standard techniques. The breaking of rotational symmetry due to the presence of a catalysing defect complicates the equations which determine the field profiles describing the critical bubble. To overcome these difficulties we use the mountain pass algorithm, which is well-suited for finding saddle points in phase transitions with reduced symmetry.

One of the parameters that is highly relevant for the phenomenology of the catalysed transition is the number of domain walls per Hubble volume $\xi$ at the onset of the
catalysed transition. For sparse networks with $\xi \lesssim \mathcal{O}(10)$ the bubble size is controlled by the correlation length of the network, leading to larger bubbles and greater gravitational wave amplitude. On the other hand, for dense networks with $\xi \gtrsim \mathcal{O}(100)$ the catalysed transition is expected to resemble a homogeneous process.
It will be very interesting to study the domain wall network explicitly in this model through numerical simulations, possibly taking into account friction effects, in order to predict the possible values of $\xi$.

Throughout our analysis, the breaking of rotational symmetry due to the presence of the impurities has had only indirect consequences on the GW spectrum (namely by setting the relevant thermodynamical quantities). However, few open questions remain on this matter. For instance, it is not yet conclusive whether the elliptical shape of the critical bubbles nucleated on the walls will be maintained in the growth to macroscopic size, even though the approach to spherical shape is the most likely outcome. In addition, for sparse networks seeding a phase transition with a runaway behaviour, the acceleration of planar walls (as opposed to spherical bubbles) may contribute to the GW spectrum in a significant way also before collision.

In summary, gravitational wave signatures of early universe phase transitions are
a promising future probe of fundamental physics. 
We have shown that just like everyday phase transitions, it is possible that the
EWPT is dominated by defects which are crucial for the nucleation of bubbles. This mechanism is very general and applicable to other cosmological phase transitions as well, where defects and impurities can play a decisive role in determining the corresponding phenomenology.

\begin{acknowledgments}
We thank Ed Copeland, Thomas Konstandin, John March-Russell  
and Diego Redigolo for useful discussions and comments.
P.A. is supported by the STFC under Grant No. ST/T000864/1. MN is supported in part by a joint Clarendon and Sloane-Robinson scholarship from Oxford University and Keble college, and in part by the National Science Foundation under Grant No. PHY-2310717.
SB and AM are supported in part by the Strategic Research
Program High-Energy Physics of the Research Council
of the Vrije Universiteit Brussel and by the iBOF ``Unlocking the Dark Universe with Gravitational Wave Observations: from Quantum Optics to Quantum Gravity'' of the Vlaamse Interuniversitaire Raad. SB and AM are also supported in part by the ``Excellence
of Science - EOS'' - be.h project n.30820817.
SB is supported in part by FWO-Vlaanderen through grant numbers 12B2323N.
This work is supported by the Deutsche Forschungsgemeinschaft under Germany’s Excellence Strategy - EXC 2121 Quantum Universe - 390833306.
\end{acknowledgments}

\appendix

\section{Comparison to previous work}

\label{sec:highT}

In this appendix, we provide a comparison between the results obtained
within the domain wall effective field theory (EFT) and the MPT algorithm by
retaining only the leading terms in the high--temperature
approximation. This provides a non trivial cross check of our methods
and corroborates our strategy in view of generalising these results to
full 1-loop thermal potentials. 

\begin{figure}[t!]
\centering
 \includegraphics[scale=0.34]{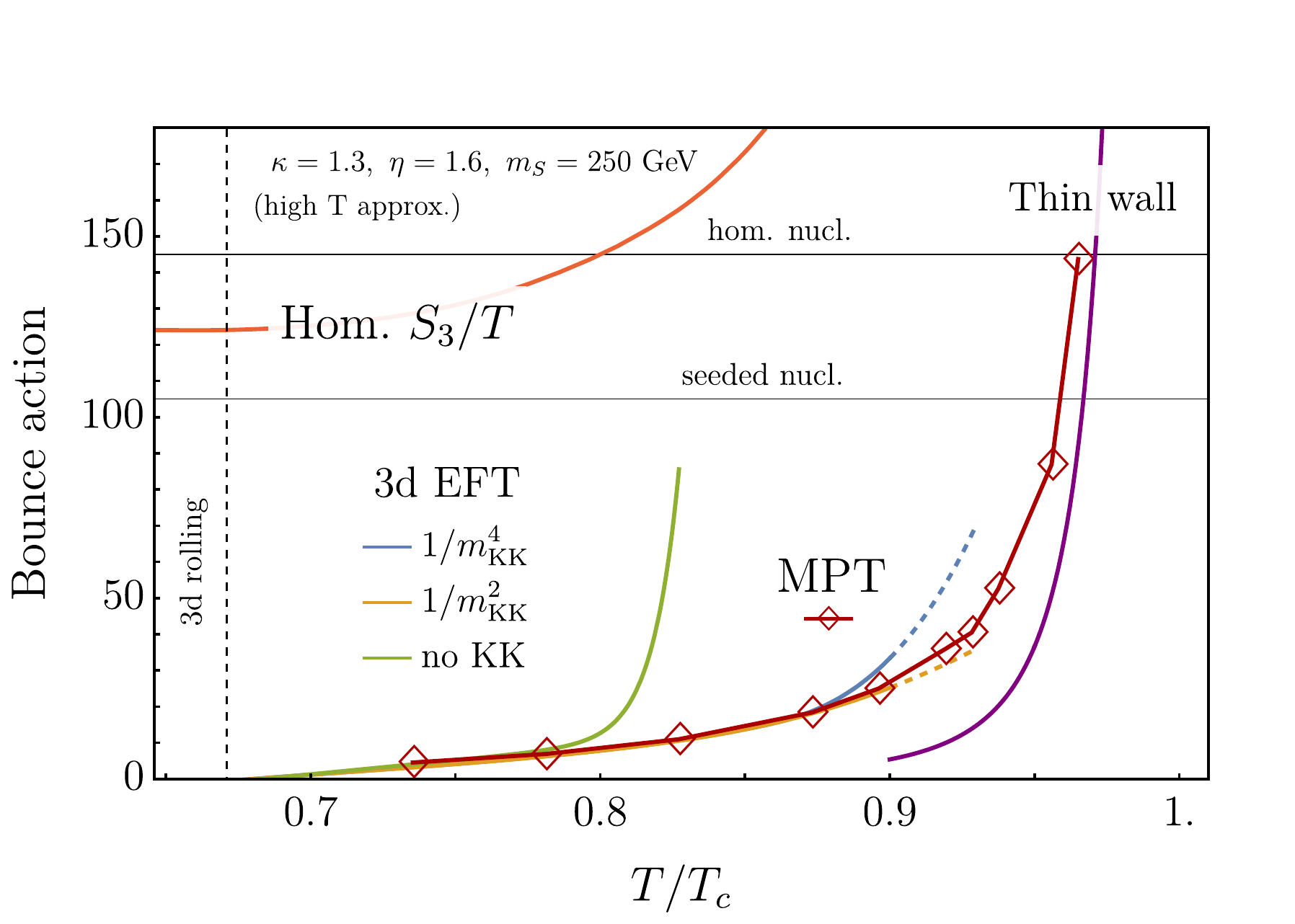}
 \caption{Comparison of the homogeneous bounce action $S_3/T$ (red) in the leading high--temperature approximation, and the seeded bounce action evaluated with the MPT (red diamonds) and within the EFT within different approximations: zeroth order where KK states are neglected in green, $\mathcal{O}(1/m_{\rm KK}^2)$ in orange and $\mathcal{O}(1/m_{\rm KK}^4)$ in blue. The purple line shows the seeded bounce action within the thin wall approximation.}
\label{fig:warmup} 
\end{figure}

In figure~\ref{fig:warmup} we compare the tunneling action evaluated with the domain wall EFT (described in further detail in ref~\cite{Blasi:2022woz}) and the MPT algorithm, for the benchmark point given by
$\kappa = 1.3$, $\eta = 1.6$ and $\mu_s \simeq 127\,\text{GeV}$ leading to a singlet mass in the true vacuum $m_S = 250 \,\text{GeV}$ at zero temperature. For this benchmark the critical temperature is $T_c \simeq 110 \,\text{GeV}$. At 
$T_r \simeq 74 \,\text{GeV}$ the tunnelling action approaches zero, signaling a classical instability of the domain walls.

The temperature range where the EFT is supposed to provide reliable results for the bounce action can be estimated by considering the ratio between the lightest Higgs zero mode mass, $\omega_0^2(T)$ (see equation~\eqref{eq:omega_0}),
and the mass scale of the continuum KK states, $m^2_\text{KK}(T)$.
When this ratio is small, integrating out the KK states is indeed justified and the expansion in terms of the inverse KK mass is supposed to be converging.
In practice, it is more convenient to identify the range of validity by comparing the prediction for the bounce action keeping different $1/m_{\rm KK}^2$ orders in the EFT. As we can see from figure~\ref{fig:warmup}, the EFT works very well at lower temperatures close to the 3d rolling, where in fact $\omega_0^2(T \approx T_r) \approx 0$ and a large hierarchy exists between the lightest Higgs mode and the KK states. The agreement between the different predictions degrades for higher temperatures, and for $T/T_c \gtrsim 0.9$ the $1/m_{\rm KK}^2$ expansion can be considered broken. This is shown in figure~\ref{fig:warmup} by the dashed lines. The high temperature region is in fact difficult to treat with the domain wall EFT, as the seeded bubble tends to extend further in the direction orthogonal to the wall and this behaviour can only be captured by including more and more KK states.

As shown in ref.\,\cite{Blasi:2022woz}, for temperatures close to $T_c$ one can still rely on a thin wall approximation for the seeded tunneling similarly to the homogenous case. Further details and modifications compared to ref.\,\cite{Blasi:2022woz} are discussed in appendix~\ref{app:ThinWall}. This calculation is limited by the fact that away from $T\approx T_c$ the bubble wall becomes thick and this approximation is no longer valid.
The results from the thin wall limit are shown in figure~\ref{fig:warmup} by the purple line until $T = 0.9 T_c$. As detailed in appendix~\ref{app:ThinWall}, the definition of our thin wall implies that the resulting bounce action is actually a lower bound on the true one.

Considering the limitation of the thin wall at low temperatures, and of the EFT at high temperatures, there remains a gap in which the tunneling rate is not calculable. We can however overcome these limitations by calculating the bounce action fully numerically via the MPT algorithm presented in the main text. The results are shown by the red diamonds in figure~\ref{fig:warmup}. The MPT calculation of the bounce action matches very precisely the EFT at low temperatures, as well as the thin wall at high temperatures (consistently with the thin wall providing a lower bound, as mentioned above), hence the MPT provides the correct extrapolation in the intermediate temperature range.

\section{The thin wall approximation}
\label{app:ThinWall}

In this Appendix we review how to compute the bounce action for the seeded decay in the thin tall approximation
\cite{Blasi:2022woz}.
We first discuss the thin wall approximation to compute the bounce action for the homogeneous decay of the false vacuum
in a generic two field model. Then we extend the same strategy to the false vacuum decay seeded by the domain walls.

\paragraph{Thin wall for homogeneous bounce in a two--field model.}
We would like to estimate the $O(3)$ symmetric bounce action in the thin wall (TW) regime for a model with two fields.
The equation of motions for the two fields are (notation $\phi_i={h,S}$)
\be
\label{eommulti}
\frac{d^2 \phi_i}{dr^2} +\frac{2}{r}\frac{d \phi_i}{dr} = \frac{\partial V}{ \partial \phi_i}
\ee
We denote with $\phi_i^T$ and $\phi_i^F$ the true and false vacuum respectively.
The boundary conditions are
\be
\label{bdry}
\phi_i(r\to \infty) \to \phi_i^F \qquad \phi_i'(0)=0.
\ee
We define
$
\Delta V = V(\phi_i^F)-V(\phi_i^T) >0
$
as the vacuum difference.

The bounce action is the action evaluated on the bubble solution, minus the action evaluated in the false vacuum, where
\be
S_{\text{Bubble}} = 4 \pi \int r^2 dr \left(\frac{1}{2} \sum_i \left(\frac{d \phi_i}{dr} \right)^2 \right) + V,
\quad 
S_{\text{false}} = 4\pi \int r^2 dr V.
\ee
The TW approximation captures the value of the bounce action in the limit where the two minima are almost degenerate $\Delta V \to 0$
\cite{Coleman:1977py,Anderson:1991zb}.
In this limit, the energetically favourable field configuration is such that the fields vary significantly only at the bubble wall, in a region 
$\delta R$ which is much smaller than the bubble radius $R$.
In the region $r>R+\delta R$ the fields are in the false vacuum, while in the region 
$r<R-\delta R$ they sit on release points $\phi_i^* $ which lie between the false and the true vacuum.
For this configuration it is a good approximation to neglect the friction term in the equation \eqref{eommulti}.
Working at leading order in an expansion $\delta R \ll R$ the bounce action is
\be
\label{bounce_TW_hom}
S_{\text{bounce}}  \simeq 4 \pi  R^2 \sigma_{\rm B} -\frac{4}{3} \pi R^3 \Delta V,
\ee
where the bubble tension in the TW approximation is defined as
\be
\sigma_{\rm B}=
\int_{R - \delta R}^{R+\delta R}dr \left(\frac{1}{2} \sum_i \left(\frac{d \phi_i}{dr} \right)^2 +(V-V(\phi_i^*)) \right).
\ee
Minimizing $S_{\text{bounce}}$  with respect to the radius, one finds the usual formula
\cite{Coleman:1977py,Linde:1981zj}
\be
S_\text{bounce}= \frac{16 \pi \sigma_B^3}{3 \Delta V^2}.
\ee

The non-trivial step is to compute the bubble tension within the TW approximation in the two field model.
By employing the equation of motions in the TW approximation, together with the boundary conditions \eqref{bdry},
we obtain
\be
\label{relation}
\sum_i \left( \frac{ d \phi_i}{dr} \right)^2   = 2 (V - V(\phi_i^*) ).
\ee
With this equation we conclude that the tension is simply given by
\be
\sigma_{\rm B}= \int_{R - \delta R}^{R+\delta R}dr \left( 2 (V - V(\phi_i^*) )\right).
\ee
We then
change integration variable in the previous expression by using again \eqref{relation}.
We specialize to two fields.
The field $\phi_1$ goes from $\phi_1(0) = \phi_1^*$ to $\phi_1(\infty) = \phi_1^F$. Given that $\phi_1$ is the Higgs field, the first derivative of $\phi_1$ with respect to $r$
is expected to be negative.
We parametrize the path in the field space with $\phi_2$ as a function of $\phi_1$ such that $\phi_2(\phi_1^*) = \phi_2^*$ and $\phi_2(\phi_1^F)=\phi_2^F$.
We can then use \eqref{relation} to show that
\be
\frac{d \phi_1}{dr} =- \frac{\sqrt{2 (V - V(\phi_i^*) )}}{\sqrt{1+\left(\frac{d \phi_2}{d\phi_1}\right)^2}},
\ee
where we choose the negative branch because of the previous argument.
Given that $\phi_1(R+\delta R) =\phi_1^F$ and $\phi_1(R-\delta R)=\phi_1^*$,
we conclude that 
\be
\label{final}
\sigma_{\rm B} = \int_{\phi_1^F}^{\phi_1^*}  \sqrt{1+\left(\frac{d \phi_2}{d\phi_1}\right)^2} \sqrt{2 (V - V(\phi_i^*) )}.
\ee

The minimal value for the tension is obtained when the integral extends the least on the $\phi_1$ range, namely for $\phi_1^* = \phi_1^c$ where $\phi_1^c$ is defined as the point in the field path on the descending
slope of the potential where $V(\phi_1^c) = V(\phi_1^F)$.
This is the prescription that we adopt in the main text as well as in appendix~\ref{sec:highT}, implying that our thin wall results are supposed to be a lower bound on the true bounce action (also for the seeded tunneling, as discussed below).

\paragraph{Thin wall for the seeded bounce action.}
The $O(3)$ bounce action at finite temperature can be interpreted 
as the energy difference between the bubble of radius $R$ and the false vacuum \cite{Linde:1981zj}.
In the presence of the DW, this difference involves an extra piece with respect to equation \eqref{bounce_TW_hom}
corresponding to the fact that a portion of the original DW is removed inside the bubble\,\cite{Blasi:2022woz}:
\be
\label{bounce_TW_DW}
S_{\text{bounce}}  \simeq 4 \pi  R^2 \sigma_{\rm B} -\frac{4}{3} \pi R^3 \Delta V - 4\pi R^2 \sigma_{\rm DW}
\ee
where $\sigma_{\rm DW}$ is the DW tension.
In deriving \eqref{bounce_TW_DW} we assumed $O(3)$ 
symmetry so that the portion of DW removed in the tunneling is a disk of radius $R$, and also that the bubble tension $\sigma_{\rm B}$ in the seeded bounce is
well approximated by the bubble tension of the homogeneous bubble in \eqref{final}. These are robust assumptions far from the DW, so we expect \eqref{bounce_TW_DW} to be a good approximation if the bubble radius is much larger than the DW width\,\footnote{In \cite{Blasi:2022woz} the possibility of a $O(2)$ symmetric TW bounce was considered. However, this possibility did not lead to
significant changes in the bounce estimate, and hence here we stick to the $O(3)$ case.}.
By minimizing \eqref{bounce_TW_DW} with respect to the bubble radius, we
derive
\be
\label{eq:finalDW}
S_{\text{bounce}}^{\text{DW}} = 
\frac{16 \pi \left( \sigma_{\rm B} - \sigma_{\rm DW}/4\right)^3}{3 \Delta V^2}
\ee
which is the formula of \cite{Blasi:2022woz} and the one employed in our analysis.
Note that in order to evaluate $\sigma_{\rm B}$ in \eqref{eq:finalDW} according to its expression in \eqref{final}, one only needs to know the path in field space for the homogeneous tunneling solution. We have used the public code \texttt{Findbounce} \cite{Guada:2020xnz} to obtain the path profile and compute the bubble tension with the prescription discussed below \eqref{final}, so that \eqref{eq:finalDW} can be considered a lower bound on the seeded bounce action.

\section{Percolation in the presence of monopoles, strings and walls}
\label{sec:percgen}

In this appendix we collect some results regarding percolation in the presence of defects of various codimensions. This analysis applies to sparse networks in which the distance among the defects sets the time scale for percolation, so that we can assume that at the same nucleation time $t_\nuc$ all the defects become unstable and start expanding converting the space to the true vacuum. 

Our strategy is to generalize the expression for percolation valid for homogeneous phase transitions to the case of impurities. We start from the definition\,\cite{Guth:1981uk} adapted to a general $D$--dimensional space:
\be
\label{eq:generalI}
I_D (t) \equiv \int_{t_c}^t \text{d} t^\prime \Gamma(t^\prime) \,
a^D(t^\prime) V(t,t^\prime),
\ee
where $V(t,t^\prime)$ is the $D$--dimensional co-moving volume converted to the true vacuum at the time $t$ considering a nucleation event at the time $t^\prime$. $\Gamma(t)$ is as usual the nucleation rate per unit time per unit (physical) volume. The integral in \eqref{eq:generalI} is related to the probability of a point being in the false vacuum at the time $t$ as $p(t) = e^{-I_D(t)}$. Eq.\,\eqref{eq:3Dperc} is related to \eqref{eq:generalI} by assuming the nucleation of spherical bubbles in a 3D space.

\medskip

Let us now apply \eqref{eq:generalI} to the case of monopoles acting as nucleation sites\,\footnote{The original result can be found in ref.\,\cite{Guth:1981uk}.}. We then consider a sharply peaked nucleation rate around the monopoles at the time $t_\nuc$ (which could be interpreted as the time when their cores become unstable), and a finite density of monopoles $n_{\rm m}(t)$. This leads to the following nucleation rate per unit volume:
\be
\Gamma_3(t) = \delta(t - t_\nuc) n_{\rm m}(t) = \delta(t - t_\nuc) \xi_{\rm m}(t) \, H^{3}(t),
\ee
where we have introduced the parameter $\xi_{\rm m}(t)$ as the number of monopoles per Hubble volume, including a possible time dependence encoding the dynamics of the network, and the suffix indicates the codimension $D=3$ of monopoles.
Plugging this back in \eqref{eq:generalI} one obtains
\be
\label{eq:percmon}
I_{\rm m}(t) =  \xi_{\rm m}(t_n) H(t_\nuc)^3 \cdot \frac{4\pi}{3}
\left[ \int_{t_\nuc}^t v_w \frac{a(t_\nuc)}{a(\tilde t)}  \,{\rm d} \tilde t \right]^3,
\ee
where we have used that bubbles around monopoles are 3D spheres. This result agrees with Eq.\,(7.3) in ref.\,\cite{Guth:1981uk}.

\medskip

In order to apply the same analysis to a network of strings or walls, one has to take into account the different codimension of these defects.
Let us first consider the case of strings, whose codimension is $D=2$, and idealize the network as a collection of straight, infinite strings parallel to the $z$ axis. 
The number of such strings per Hubble volume will be denoted by $\xi_{\rm s}(t)$.
When nucleation occurs, the string core starts to expand in the radial direction at every point along the string. Considering a slice of space $z = z_0$ orthogonal to the strings, the progress of the transition can be measured by looking at the disks of true vacuum that are expanding from the points where the strings pierce our two--dimensional slice. The number of such points is determined by $\xi_{\rm s}(t_\nuc)$, and we can therefore consider the following nucleation rate per unit surface in 2D:
\be
\Gamma_2(t) = \delta(t - t_\nuc) \, \xi_{\rm s}(t) H^2(t),
\ee
where the factor $\xi_{\rm s}(t) H^2(t)$ stands for the number of string--crossing points per unit surface. Substituting this back in \eqref{eq:generalI} with $D=2$ one obtains:
\be
I_{\rm s}(t) = \xi_{\rm s}(t_\nuc) H^2(t_\nuc) \cdot \pi \left[ \int_{t_\nuc}^t v_w \frac{a(t_\nuc)}{a(\tilde t)}  \,{\rm d} \tilde t \right]^2.
\ee

\medskip

Finally, let us apply the same reasoning to a network of domain walls seeding the phase transition, whose codimension is $D=1$. Consider a network of walls parallel to the $x$--$y$ plane. The number of walls in one Hubble volume will be indicated by $\xi_{\rm w}(t)$ as usual. The progress of the transition can be measured by considering straight lines orthogonal to the walls (parallel to the $z$ axis in this case). Such straight lines intersect the walls in a given number of points depending on $\xi_{\rm w}(t)$, from which the conversion to the true vacuum begins. Along the lines orthogonal to the walls, the transition proceeds with nucleation rate given by
\be
\Gamma_1(t) = \delta(t - t_\nuc) \, \xi_{\rm w}(t) H(t).
\ee
Referring to \eqref{eq:generalI} with $D=1$, one then finds
\be
I_{\rm w}(t) = \xi_{\rm w}(t_\nuc) H(t_\nuc) \cdot 2 \int_{t_\nuc}^t v_w \frac{a(t_\nuc)}{a(\tilde t)}  \,{\rm d} \tilde t,
\ee
which corresponds to Eq.\,\eqref{eq:percseed} presented in the main text.

\medskip

We can summarize our results for the probability of a point to be in the false vacuum during a phase transition seeded by a sparse network of defects with codimension $D$ as
\be
\label{eq:IDt}
p(t) = e^{-I_D(t)}, \quad I_D(t) = \xi(t_\nuc) \,H^D(t_\nuc) \frac{\pi^{D/2}}{\Gamma\left( \frac{D}{2}+1 \right)} \left[a(t_\nuc)\, r(t,t_\nuc)\right]^D,
\ee
where $D=1,2,3$ corresponds to domain walls, cosmic strings, and monopoles, respectively, $\xi$ indicates the number of defects per Hubble patch ($\xi \sim 1$ for sparse networks), and $r(t,t^\prime)$ is given in \eqref{eq:radius}. Notice that $I_D(t)$ depends on the time $t$ only through $r(t, t_\nuc)$.

\medskip 

Given \eqref{eq:IDt} one can define the time of percolation as
$I_D(t_\perc) = n_c$, where $n_c$ is the percolation threshold. In the main text we use $n_c= 0.34$ both for homogeneous and domain--wall seeded phase transitions.

\medskip

Let us finally discuss the condition for the decrease of the physical volume in the false vacuum as the phase transition progresses, which is given by
\be
V_{\rm false}(t) \propto a^3(t) \, e^{-I_D(t)}.
\ee
Notice that, independently of the defect codimension, the volume in the false vacuum still increases as $\propto a^3$, since this only depends on the embedding 3D space. The condition for volume decrease then simply reads
\be
\label{eq:volumedecreaseD}
\frac{1}{V_{\rm false}} \frac{\text{d}}{\text{d}t} V_{\rm false} = 
3 H(t) -  \frac{d I_D(t)}{d t} < 0.
\ee 

We can evaluate the condition above analytically by considering the universe to be radiation dominated at percolation, defined for all codimesions as $I_D(t_\perc)=n_c$. 
In the case of monopoles, or in general point--like defects, we find that
\be
	3H_\perc - \frac{d I_{\rm m}(t)}{d t}\bigg|_{t=t_\perc} = 
	H_\perc \left(  3 - 3 n_c - 3 v_w \left( \frac{4\pi}{3} n_c^2 \,\xi_{\rm m} \right)^{1/3}  \right) < 0.
\ee
For strings, one has 
\be
	3H_\perc - \frac{d I_{\rm s}(t)}{d t}\bigg|_{t=t_\perc} = 
	H_\perc \left( 3 - 2 n_c - 2 v_w \left(\pi \, n_c \, \xi_{\rm s}\right)^{1/2}  \right)<0,
\ee
and finally for domain walls
\be
	3H_\perc - \frac{d I_{\rm w}(t)}{d t}\bigg|_{t=t_\perc} = 
	H_\perc \left(  3 - n_c - 2 v_w \xi_{\rm w}  \right) <0,
\ee
as presented in the main text.

\bibliographystyle{JHEP}
\bibliography{bibfr}

\providecommand{\href}[2]{#2}\begingroup\raggedright\begin{thebibliography}{100}

\bibitem{LISA:2017pwj}
{\bf LISA}, P.~Amaro-Seoane et~al., {\it {Laser Interferometer Space Antenna}},
   \href{http://arxiv.org/abs/1702.00786}{{\tt arXiv:1702.00786}}.

\bibitem{Caprini:2015zlo}
C.~Caprini et~al., {\it {Science with the space-based interferometer eLISA. II:
  Gravitational waves from cosmological phase transitions}},  {\em JCAP} {\bf
  04} (2016) 001, [\href{http://arxiv.org/abs/1512.06239}{{\tt
  arXiv:1512.06239}}].

\bibitem{Caprini:2019egz}
C.~Caprini et~al., {\it {Detecting gravitational waves from cosmological phase
  transitions with LISA: an update}},  {\em JCAP} {\bf 03} (2020) 024,
  [\href{http://arxiv.org/abs/1910.13125}{{\tt arXiv:1910.13125}}].

\bibitem{Kajantie:1995kf}
K.~Kajantie, M.~Laine, K.~Rummukainen, and M.~E. Shaposhnikov, {\it {The
  Electroweak phase transition: A Nonperturbative analysis}},  {\em Nucl. Phys.
  B} {\bf 466} (1996) 189--258,
  [\href{http://arxiv.org/abs/hep-lat/9510020}{{\tt hep-lat/9510020}}].

\bibitem{Kajantie:1996mn}
K.~Kajantie, M.~Laine, K.~Rummukainen, and M.~E. Shaposhnikov, {\it {Is there
  a~ hot electroweak phase transition at $m_H \gtrsim m_W$?}},  {\em Phys. Rev.
  Lett.} {\bf 77} (1996) 2887--2890,
  [\href{http://arxiv.org/abs/hep-ph/9605288}{{\tt hep-ph/9605288}}].

\bibitem{Karsch:1996yh}
F.~Karsch, T.~Neuhaus, A.~Patkos, and J.~Rank, {\it {Critical Higgs mass and
  temperature dependence of gauge boson masses in the SU(2) gauge Higgs
  model}},  {\em Nucl. Phys. B Proc. Suppl.} {\bf 53} (1997) 623--625,
  [\href{http://arxiv.org/abs/hep-lat/9608087}{{\tt hep-lat/9608087}}].

\bibitem{Aoki:1996pn}
Y.~Aoki, {\it {Four-dimensional simulation of the hot electroweak phase
  transition with the SU(2) gauge Higgs model}},  {\em Nucl. Phys. B Proc.
  Suppl.} {\bf 53} (1997) 609--611,
  [\href{http://arxiv.org/abs/hep-lat/9608061}{{\tt hep-lat/9608061}}].

\bibitem{Gurtler:1997hr}
M.~Gurtler, E.-M. Ilgenfritz, and A.~Schiller, {\it {Where the electroweak
  phase transition ends}},  {\em Phys. Rev. D} {\bf 56} (1997) 3888--3895,
  [\href{http://arxiv.org/abs/hep-lat/9704013}{{\tt hep-lat/9704013}}].

\bibitem{Laine:1998jb}
M.~Laine and K.~Rummukainen, {\it {What's new with the electroweak phase
  transition?}},  {\em Nucl. Phys. B Proc. Suppl.} {\bf 73} (1999) 180--185,
  [\href{http://arxiv.org/abs/hep-lat/9809045}{{\tt hep-lat/9809045}}].

\bibitem{Pietroni:1992in}
M.~Pietroni, {\it {The Electroweak phase transition in a nonminimal
  supersymmetric model}},  {\em Nucl. Phys. B} {\bf 402} (1993) 27--45,
  [\href{http://arxiv.org/abs/hep-ph/9207227}{{\tt hep-ph/9207227}}].

\bibitem{Carena:1996wj}
M.~Carena, M.~Quiros, and C.~E.~M. Wagner, {\it {Opening the window for
  electroweak baryogenesis}},  {\em Phys. Lett. B} {\bf 380} (1996) 81--91,
  [\href{http://arxiv.org/abs/hep-ph/9603420}{{\tt hep-ph/9603420}}].

\bibitem{Delepine:1996vn}
D.~Delepine, J.~M. Gerard, R.~Gonzalez~Felipe, and J.~Weyers, {\it {A Light
  stop and electroweak baryogenesis}},  {\em Phys. Lett. B} {\bf 386} (1996)
  183--188, [\href{http://arxiv.org/abs/hep-ph/9604440}{{\tt hep-ph/9604440}}].

\bibitem{Apreda:2001us}
R.~Apreda, M.~Maggiore, A.~Nicolis, and A.~Riotto, {\it {Gravitational waves
  from electroweak phase transitions}},  {\em Nucl. Phys. B} {\bf 631} (2002)
  342--368, [\href{http://arxiv.org/abs/gr-qc/0107033}{{\tt gr-qc/0107033}}].

\bibitem{Huber:2015znp}
S.~J. Huber, T.~Konstandin, G.~Nardini, and I.~Rues, {\it {Detectable
  Gravitational Waves from Very Strong Phase Transitions in the General
  NMSSM}},  {\em JCAP} {\bf 03} (2016) 036,
  [\href{http://arxiv.org/abs/1512.06357}{{\tt arXiv:1512.06357}}].

\bibitem{Kuzmin:1985mm}
V.~A. Kuzmin, V.~A. Rubakov, and M.~E. Shaposhnikov, {\it {On the Anomalous
  Electroweak Baryon Number Nonconservation in the Early Universe}},  {\em
  Phys. Lett. B} {\bf 155} (1985) 36.

\bibitem{Shaposhnikov:1987tw}
M.~E. Shaposhnikov, {\it {Baryon Asymmetry of the Universe in Standard
  Electroweak Theory}},  {\em Nucl. Phys. B} {\bf 287} (1987) 757--775.

\bibitem{Rubakov:1996vz}
V.~A. Rubakov and M.~E. Shaposhnikov, {\it {Electroweak baryon number
  nonconservation in the early universe and in high-energy collisions}},  {\em
  Usp. Fiz. Nauk} {\bf 166} (1996) 493--537,
  [\href{http://arxiv.org/abs/hep-ph/9603208}{{\tt hep-ph/9603208}}].

\bibitem{Kosowsky:1991ua}
A.~Kosowsky, M.~S. Turner, and R.~Watkins, {\it {Gravitational radiation from
  colliding vacuum bubbles}},  {\em Phys. Rev. D} {\bf 45} (1992) 4514--4535.

\bibitem{Kosowsky:1992rz}
A.~Kosowsky, M.~S. Turner, and R.~Watkins, {\it {Gravitational waves from first
  order cosmological phase transitions}},  {\em Phys. Rev. Lett.} {\bf 69}
  (1992) 2026--2029.

\bibitem{Kosowsky:1992vn}
A.~Kosowsky and M.~S. Turner, {\it {Gravitational radiation from colliding
  vacuum bubbles: envelope approximation to many bubble collisions}},  {\em
  Phys. Rev. D} {\bf 47} (1993) 4372--4391,
  [\href{http://arxiv.org/abs/astro-ph/9211004}{{\tt astro-ph/9211004}}].

\bibitem{Kamionkowski:1993fg}
M.~Kamionkowski, A.~Kosowsky, and M.~S. Turner, {\it {Gravitational radiation
  from first order phase transitions}},  {\em Phys. Rev. D} {\bf 49} (1994)
  2837--2851, [\href{http://arxiv.org/abs/astro-ph/9310044}{{\tt
  astro-ph/9310044}}].

\bibitem{Athron:2023xlk}
P.~Athron, C.~Bal\'azs, A.~Fowlie, L.~Morris, and L.~Wu, {\it {Cosmological
  phase transitions: From perturbative particle physics to gravitational
  waves}},  {\em Prog. Part. Nucl. Phys.} {\bf 135} (2024) 104094,
  [\href{http://arxiv.org/abs/2305.02357}{{\tt arXiv:2305.02357}}].

\bibitem{Hashino:2016rvx}
K.~Hashino, M.~Kakizaki, S.~Kanemura, and T.~Matsui, {\it {Synergy between
  measurements of gravitational waves and the triple-Higgs coupling in probing
  the first-order electroweak phase transition}},  {\em Phys. Rev. D} {\bf 94}
  (2016), no.~1 015005, [\href{http://arxiv.org/abs/1604.02069}{{\tt
  arXiv:1604.02069}}].

\bibitem{Huang:2016cjm}
P.~Huang, A.~J. Long, and L.-T. Wang, {\it {Probing the Electroweak Phase
  Transition with Higgs Factories and Gravitational Waves}},  {\em Phys. Rev.
  D} {\bf 94} (2016), no.~7 075008,
  [\href{http://arxiv.org/abs/1608.06619}{{\tt arXiv:1608.06619}}].

\bibitem{Hashino:2016xoj}
K.~Hashino, M.~Kakizaki, S.~Kanemura, P.~Ko, and T.~Matsui, {\it {Gravitational
  waves and Higgs boson couplings for exploring first order phase transition in
  the model with a singlet scalar field}},  {\em Phys. Lett. B} {\bf 766}
  (2017) 49--54, [\href{http://arxiv.org/abs/1609.00297}{{\tt
  arXiv:1609.00297}}].

\bibitem{Artymowski:2016tme}
M.~Artymowski, M.~Lewicki, and J.~D. Wells, {\it {Gravitational wave and
  collider implications of electroweak baryogenesis aided by non-standard
  cosmology}},  {\em JHEP} {\bf 03} (2017) 066,
  [\href{http://arxiv.org/abs/1609.07143}{{\tt arXiv:1609.07143}}].

\bibitem{Beniwal:2017eik}
A.~Beniwal, M.~Lewicki, J.~D. Wells, M.~White, and A.~G. Williams, {\it
  {Gravitational wave, collider and dark matter signals from a scalar singlet
  electroweak baryogenesis}},  {\em JHEP} {\bf 08} (2017) 108,
  [\href{http://arxiv.org/abs/1702.06124}{{\tt arXiv:1702.06124}}].

\bibitem{Coleman:1977py}
S.~R. Coleman, {\it {The Fate of the False Vacuum. 1. Semiclassical Theory}},
  {\em Phys. Rev. D} {\bf 15} (1977) 2929--2936. [Erratum: Phys.Rev.D 16, 1248
  (1977)].

\bibitem{Callan:1977pt}
C.~G. Callan, Jr. and S.~R. ~, {\it {The Fate of the False Vacuum. 2. First
  Quantum Corrections}},  {\em Phys. Rev. D} {\bf 16} (1977) 1762--1768.

\bibitem{Linde:1981zj}
A.~D. Linde, {\it {Decay of the False Vacuum at Finite Temperature}},  {\em
  Nucl. Phys. B} {\bf 216} (1983) 421. [Erratum: Nucl.Phys.B 223, 544 (1983)].

\bibitem{Steinhardt:1981mm}
P.~J. Steinhardt, {\it {Monopole Dissociation in the Early Universe}},  {\em
  Phys. Rev. D} {\bf 24} (1981) 842.

\bibitem{Steinhardt:1981ec}
P.~J. Steinhardt, {\it {Monopole and Vortex Dissociation and Decay of the False
  Vacuum}},  {\em Nucl. Phys. B} {\bf 190} (1981) 583--616.

\bibitem{Hosotani:1982ii}
Y.~Hosotani, {\it {Impurities in the Early Universe}},  {\em Phys. Rev. D} {\bf
  27} (1983) 789.

\bibitem{Jensen:1982jv}
L.~G. Jensen and P.~J. Steinhardt, {\it {DISSOCIATION OF
  ABRIKOSOV-NIELSEN-OLESEN VORTICES}},  {\em Phys. Rev. B} {\bf 27} (1983)
  5549.

\bibitem{Witten:1984rs}
E.~Witten, {\it {Cosmic Separation of Phases}},  {\em Phys. Rev. D} {\bf 30}
  (1984) 272--285.

\bibitem{Yajnik:1986tg}
U.~A. Yajnik, {\it {PHASE TRANSITION INDUCED BY COSMIC STRINGS}},  {\em Phys.
  Rev. D} {\bf 34} (1986) 1237--1240.

\bibitem{Yajnik:1986wq}
U.~A. Yajnik and T.~Padmanabhan, {\it {ANALYTICAL APPROACH TO STRING INDUCED
  PHASE TRANSITION}},  {\em Phys. Rev. D} {\bf 35} (1987) 3100.

\bibitem{Hiscock:1987hn}
W.~A. Hiscock, {\it {CAN BLACK HOLES NUCLEATE VACUUM PHASE TRANSITIONS?}},
  {\em Phys. Rev. D} {\bf 35} (1987) 1161--1170.

\bibitem{Berezin:1987ea}
V.~A. Berezin, V.~A. Kuzmin, and I.~I. Tkachev, {\it {O(3) Invariant Tunneling
  in General Relativity}},  {\em Phys. Lett. B} {\bf 207} (1988) 397--403.

\bibitem{Arnold:1989cq}
P.~B. Arnold, {\it {GRAVITY AND FALSE VACUUM DECAY RATES: O(3) SOLUTIONS}},
  {\em Nucl. Phys. B} {\bf 346} (1990) 160--192.

\bibitem{Berezin:1990qs}
V.~A. Berezin, V.~A. Kuzmin, and I.~I. Tkachev, {\it {Black holes initiate
  false vacuum decay}},  {\em Phys. Rev. D} {\bf 43} (1991) 3112--3116.

\bibitem{Preskill:1992ck}
J.~Preskill and A.~Vilenkin, {\it {Decay of metastable topological defects}},
  {\em Phys. Rev. D} {\bf 47} (1993) 2324--2342,
  [\href{http://arxiv.org/abs/hep-ph/9209210}{{\tt hep-ph/9209210}}].

\bibitem{Kusenko:1997hj}
A.~Kusenko, {\it {Phase transitions precipitated by solitosynthesis}},  {\em
  Phys. Lett. B} {\bf 406} (1997) 26--33,
  [\href{http://arxiv.org/abs/hep-ph/9705361}{{\tt hep-ph/9705361}}].

\bibitem{Dasgupta:1997kn}
I.~Dasgupta, {\it {Vacuum tunneling by cosmic strings}},  {\em Nucl. Phys. B}
  {\bf 506} (1997) 421--435, [\href{http://arxiv.org/abs/hep-th/9702041}{{\tt
  hep-th/9702041}}].

\bibitem{Metaxas:2000qf}
D.~Metaxas, {\it {Nontopological solitons as nucleation sites for cosmological
  phase transitions}},  {\em Phys. Rev. D} {\bf 63} (2001) 083507,
  [\href{http://arxiv.org/abs/hep-ph/0009225}{{\tt hep-ph/0009225}}].

\bibitem{Kumar:2008jb}
B.~Kumar and U.~A. Yajnik, {\it {On stability of false vacuum in supersymmetric
  theories with cosmic strings}},  {\em Phys. Rev. D} {\bf 79} (2009) 065001,
  [\href{http://arxiv.org/abs/0807.3254}{{\tt arXiv:0807.3254}}].

\bibitem{Kumar:2009pr}
B.~Kumar and U.~Yajnik, {\it {Graceful exit via monopoles in a theory with
  O'Raifeartaigh type supersymmetry breaking}},  {\em Nucl. Phys. B} {\bf 831}
  (2010) 162--177, [\href{http://arxiv.org/abs/0908.3949}{{\tt
  arXiv:0908.3949}}].

\bibitem{Kumar:2010mv}
B.~Kumar, M.~B. Paranjape, and U.~A. Yajnik, {\it {Fate of the false monopoles:
  Induced vacuum decay}},  {\em Phys. Rev. D} {\bf 82} (2010) 025022,
  [\href{http://arxiv.org/abs/1006.0693}{{\tt arXiv:1006.0693}}].

\bibitem{Pearce:2012jp}
L.~Pearce, {\it {Solitosynthesis induced phase transitions}},  {\em Phys. Rev.
  D} {\bf 85} (2012) 125022, [\href{http://arxiv.org/abs/1202.0873}{{\tt
  arXiv:1202.0873}}].

\bibitem{Lee:2013zca}
B.-H. Lee, W.~Lee, R.~MacKenzie, M.~B. Paranjape, U.~A. Yajnik, and D.-h. Yeom,
  {\it {Battle of the bulge: Decay of the thin, false cosmic string}},  {\em
  Phys. Rev. D} {\bf 88} (2013), no.~10 105008,
  [\href{http://arxiv.org/abs/1310.3005}{{\tt arXiv:1310.3005}}].

\bibitem{Gregory:2013hja}
R.~Gregory, I.~G. Moss, and B.~Withers, {\it {Black holes as bubble nucleation
  sites}},  {\em JHEP} {\bf 03} (2014) 081,
  [\href{http://arxiv.org/abs/1401.0017}{{\tt arXiv:1401.0017}}].

\bibitem{Burda:2015isa}
P.~Burda, R.~Gregory, and I.~Moss, {\it {Gravity and the stability of the Higgs
  vacuum}},  {\em Phys. Rev. Lett.} {\bf 115} (2015) 071303,
  [\href{http://arxiv.org/abs/1501.04937}{{\tt arXiv:1501.04937}}].

\bibitem{Mukaida:2017bgd}
K.~Mukaida and M.~Yamada, {\it {False Vacuum Decay Catalyzed by Black Holes}},
  {\em Phys. Rev. D} {\bf 96} (2017), no.~10 103514,
  [\href{http://arxiv.org/abs/1706.04523}{{\tt arXiv:1706.04523}}].

\bibitem{Canko:2017ebb}
D.~Canko, I.~Gialamas, G.~Jelic-Cizmek, A.~Riotto, and N.~Tetradis, {\it {On
  the Catalysis of the Electroweak Vacuum Decay by Black Holes at High
  Temperature}},  {\em Eur. Phys. J. C} {\bf 78} (2018), no.~4 328,
  [\href{http://arxiv.org/abs/1706.01364}{{\tt arXiv:1706.01364}}].

\bibitem{Kohri:2017ybt}
K.~Kohri and H.~Matsui, {\it {Electroweak Vacuum Collapse induced by Vacuum
  Fluctuations of the Higgs Field around Evaporating Black Holes}},  {\em Phys.
  Rev. D} {\bf 98} (2018), no.~12 123509,
  [\href{http://arxiv.org/abs/1708.02138}{{\tt arXiv:1708.02138}}].

\bibitem{Oshita:2018ptr}
N.~Oshita, M.~Yamada, and M.~Yamaguchi, {\it {Compact objects as the catalysts
  for vacuum decays}},  {\em Phys. Lett. B} {\bf 791} (2019) 149--155,
  [\href{http://arxiv.org/abs/1808.01382}{{\tt arXiv:1808.01382}}].

\bibitem{Jinno:2021ury}
R.~Jinno, T.~Konstandin, H.~Rubira, and J.~van~de Vis, {\it {Effect of density
  fluctuations on gravitational wave production in first-order phase
  transitions}},  {\em JCAP} {\bf 12} (2021), no.~12 019,
  [\href{http://arxiv.org/abs/2108.11947}{{\tt arXiv:2108.11947}}].

\bibitem{Shkerin:2021zbf}
A.~Shkerin and S.~Sibiryakov, {\it {Black hole induced false vacuum decay from
  first principles}},  {\em JHEP} {\bf 11} (2021) 197,
  [\href{http://arxiv.org/abs/2105.09331}{{\tt arXiv:2105.09331}}].

\bibitem{Shkerin:2021rhy}
A.~Shkerin and S.~Sibiryakov, {\it {Black hole induced false vacuum decay: the
  role of greybody factors}},  {\em JHEP} {\bf 08} (2022) 161,
  [\href{http://arxiv.org/abs/2111.08017}{{\tt arXiv:2111.08017}}].

\bibitem{Agrawal:2022hnf}
P.~Agrawal and M.~Nee, {\it {The Boring Monopole}},  {\em SciPost Phys.} {\bf
  13} (2022), no.~3 049, [\href{http://arxiv.org/abs/2202.11102}{{\tt
  arXiv:2202.11102}}].

\bibitem{Blasi:2022woz}
S.~Blasi and A.~Mariotti, {\it {Domain Walls Seeding the Electroweak Phase
  Transition}},  {\em Phys. Rev. Lett.} {\bf 129} (2022), no.~26 261303,
  [\href{http://arxiv.org/abs/2203.16450}{{\tt arXiv:2203.16450}}].

\bibitem{Briaud:2022few}
V.~Briaud, A.~Shkerin, and S.~Sibiryakov, {\it {Thermal false vacuum decay
  around black holes}},  {\em Phys. Rev. D} {\bf 106} (2022), no.~12 125001,
  [\href{http://arxiv.org/abs/2210.08028}{{\tt arXiv:2210.08028}}].

\bibitem{Jinno:2023vnr}
R.~Jinno, J.~Kume, and M.~Yamada, {\it {Super-slow phase transition catalyzed
  by BHs and the birth of baby BHs}},
  \href{http://arxiv.org/abs/2310.06901}{{\tt arXiv:2310.06901}}.

\bibitem{McDonald:1993ex}
J.~McDonald, {\it {Gauge singlet scalars as cold dark matter}},  {\em Phys.
  Rev. D} {\bf 50} (1994) 3637--3649,
  [\href{http://arxiv.org/abs/hep-ph/0702143}{{\tt hep-ph/0702143}}].

\bibitem{Burgess:2000yq}
C.~P. Burgess, M.~Pospelov, and T.~ter Veldhuis, {\it {The Minimal model of
  nonbaryonic dark matter: A Singlet scalar}},  {\em Nucl. Phys. B} {\bf 619}
  (2001) 709--728, [\href{http://arxiv.org/abs/hep-ph/0011335}{{\tt
  hep-ph/0011335}}].

\bibitem{Espinosa:2007qk}
J.~R. Espinosa and M.~Quiros, {\it {Novel Effects in Electroweak Breaking from
  a Hidden Sector}},  {\em Phys. Rev. D} {\bf 76} (2007) 076004,
  [\href{http://arxiv.org/abs/hep-ph/0701145}{{\tt hep-ph/0701145}}].

\bibitem{Profumo:2007wc}
S.~Profumo, M.~J. Ramsey-Musolf, and G.~Shaughnessy, {\it {Singlet Higgs
  phenomenology and the electroweak phase transition}},  {\em JHEP} {\bf 08}
  (2007) 010, [\href{http://arxiv.org/abs/0705.2425}{{\tt arXiv:0705.2425}}].

\bibitem{Barger:2007im}
V.~Barger, P.~Langacker, M.~McCaskey, M.~J. Ramsey-Musolf, and G.~Shaughnessy,
  {\it {LHC Phenomenology of an Extended Standard Model with a Real Scalar
  Singlet}},  {\em Phys. Rev. D} {\bf 77} (2008) 035005,
  [\href{http://arxiv.org/abs/0706.4311}{{\tt arXiv:0706.4311}}].

\bibitem{Espinosa:2008kw}
J.~R. Espinosa, T.~Konstandin, J.~M. No, and M.~Quiros, {\it {Some Cosmological
  Implications of Hidden Sectors}},  {\em Phys. Rev. D} {\bf 78} (2008) 123528,
  [\href{http://arxiv.org/abs/0809.3215}{{\tt arXiv:0809.3215}}].

\bibitem{Espinosa:2011ax}
J.~R. Espinosa, T.~Konstandin, and F.~Riva, {\it {Strong Electroweak Phase
  Transitions in the Standard Model with a Singlet}},  {\em Nucl. Phys. B} {\bf
  854} (2012) 592--630, [\href{http://arxiv.org/abs/1107.5441}{{\tt
  arXiv:1107.5441}}].

\bibitem{Cline:2012hg}
J.~M. Cline and K.~Kainulainen, {\it {Electroweak baryogenesis and dark matter
  from a singlet Higgs}},  {\em JCAP} {\bf 01} (2013) 012,
  [\href{http://arxiv.org/abs/1210.4196}{{\tt arXiv:1210.4196}}].

\bibitem{Profumo:2014opa}
S.~Profumo, M.~J. Ramsey-Musolf, C.~L. Wainwright, and P.~Winslow, {\it
  {Singlet-catalyzed electroweak phase transitions and precision Higgs boson
  studies}},  {\em Phys. Rev. D} {\bf 91} (2015), no.~3 035018,
  [\href{http://arxiv.org/abs/1407.5342}{{\tt arXiv:1407.5342}}].

\bibitem{Curtin:2014jma}
D.~Curtin, P.~Meade, and C.-T. Yu, {\it {Testing Electroweak Baryogenesis with
  Future Colliders}},  {\em JHEP} {\bf 11} (2014) 127,
  [\href{http://arxiv.org/abs/1409.0005}{{\tt arXiv:1409.0005}}].

\bibitem{Feng:2014vea}
L.~Feng, S.~Profumo, and L.~Ubaldi, {\it {Closing in on singlet scalar dark
  matter: LUX, invisible Higgs decays and gamma-ray lines}},  {\em JHEP} {\bf
  03} (2015) 045, [\href{http://arxiv.org/abs/1412.1105}{{\tt
  arXiv:1412.1105}}].

\bibitem{Craig:2014lda}
N.~Craig, H.~K. Lou, M.~McCullough, and A.~Thalapillil, {\it {The Higgs Portal
  Above Threshold}},  {\em JHEP} {\bf 02} (2016) 127,
  [\href{http://arxiv.org/abs/1412.0258}{{\tt arXiv:1412.0258}}].

\bibitem{Curtin:2016urg}
D.~Curtin, P.~Meade, and H.~Ramani, {\it {Thermal Resummation and Phase
  Transitions}},  {\em Eur. Phys. J. C} {\bf 78} (2018), no.~9 787,
  [\href{http://arxiv.org/abs/1612.00466}{{\tt arXiv:1612.00466}}].

\bibitem{Vaskonen:2016yiu}
V.~Vaskonen, {\it {Electroweak baryogenesis and gravitational waves from a real
  scalar singlet}},  {\em Phys. Rev. D} {\bf 95} (2017), no.~12 123515,
  [\href{http://arxiv.org/abs/1611.02073}{{\tt arXiv:1611.02073}}].

\bibitem{Kurup:2017dzf}
G.~Kurup and M.~Perelstein, {\it {Dynamics of Electroweak Phase Transition In
  Singlet-Scalar Extension of the Standard Model}},  {\em Phys. Rev. D} {\bf
  96} (2017), no.~1 015036, [\href{http://arxiv.org/abs/1704.03381}{{\tt
  arXiv:1704.03381}}].

\bibitem{Buttazzo:2018qqp}
D.~Buttazzo, D.~Redigolo, F.~Sala, and A.~Tesi, {\it {Fusing Vectors into
  Scalars at High Energy Lepton Colliders}},  {\em JHEP} {\bf 11} (2018) 144,
  [\href{http://arxiv.org/abs/1807.04743}{{\tt arXiv:1807.04743}}].

\bibitem{Alanne:2019bsm}
T.~Alanne, T.~Hugle, M.~Platscher, and K.~Schmitz, {\it {A fresh look at the
  gravitational-wave signal from cosmological phase transitions}},  {\em JHEP}
  {\bf 03} (2020) 004, [\href{http://arxiv.org/abs/1909.11356}{{\tt
  arXiv:1909.11356}}].

\bibitem{Carena:2019une}
M.~Carena, Z.~Liu, and Y.~Wang, {\it {Electroweak phase transition with
  spontaneous Z$_{2}$-breaking}},  {\em JHEP} {\bf 08} (2020) 107,
  [\href{http://arxiv.org/abs/1911.10206}{{\tt arXiv:1911.10206}}].

\bibitem{Schicho:2021gca}
P.~M. Schicho, T.~V.~I. Tenkanen, and J.~\"Osterman, {\it {Robust approach to
  thermal resummation: Standard Model meets a singlet}},  {\em JHEP} {\bf 06}
  (2021) 130, [\href{http://arxiv.org/abs/2102.11145}{{\tt arXiv:2102.11145}}].

\bibitem{Cline:2021iff}
J.~M. Cline, A.~Friedlander, D.-M. He, K.~Kainulainen, B.~Laurent, and
  D.~Tucker-Smith, {\it {Baryogenesis and gravity waves from a UV-completed
  electroweak phase transition}},  {\em Phys. Rev. D} {\bf 103} (2021), no.~12
  123529, [\href{http://arxiv.org/abs/2102.12490}{{\tt arXiv:2102.12490}}].

\bibitem{Laurent:2022jrs}
B.~Laurent and J.~M. Cline, {\it {First principles determination of bubble wall
  velocity}},  {\em Phys. Rev. D} {\bf 106} (2022), no.~2 023501,
  [\href{http://arxiv.org/abs/2204.13120}{{\tt arXiv:2204.13120}}].

\bibitem{Azatov:2022tii}
A.~Azatov, G.~Barni, S.~Chakraborty, M.~Vanvlasselaer, and W.~Yin, {\it
  {Ultra-relativistic bubbles from the simplest Higgs portal and their
  cosmological consequences}},  {\em JHEP} {\bf 10} (2022) 017,
  [\href{http://arxiv.org/abs/2207.02230}{{\tt arXiv:2207.02230}}].

\bibitem{Niemi:2020hto}
L.~Niemi, M.~J. Ramsey-Musolf, T.~V.~I. Tenkanen, and D.~J. Weir, {\it
  {Thermodynamics of a Two-Step Electroweak Phase Transition}},  {\em Phys.
  Rev. Lett.} {\bf 126} (2021), no.~17 171802,
  [\href{http://arxiv.org/abs/2005.11332}{{\tt arXiv:2005.11332}}].

\bibitem{Gould:2023ovu}
O.~Gould and T.~V.~I. Tenkanen, {\it {Perturbative effective field theory
  expansions for cosmological phase transitions}},
  \href{http://arxiv.org/abs/2309.01672}{{\tt arXiv:2309.01672}}.

\bibitem{Zeldovich:1974uw}
Y.~B. Zeldovich, I.~Y. Kobzarev, and L.~B. Okun, {\it {Cosmological
  Consequences of the Spontaneous Breakdown of Discrete Symmetry}},  {\em Zh.
  Eksp. Teor. Fiz.} {\bf 67} (1974) 3--11.

\bibitem{Kibble:1976sj}
T.~W.~B. Kibble, {\it {Topology of Cosmic Domains and Strings}},  {\em J. Phys.
  A} {\bf 9} (1976) 1387--1398.

\bibitem{Zurek:1985qw}
W.~H. Zurek, {\it {Cosmological Experiments in Superfluid Helium?}},  {\em
  Nature} {\bf 317} (1985) 505--508.

\bibitem{Katz:2014bha}
A.~Katz and M.~Perelstein, {\it {Higgs Couplings and Electroweak Phase
  Transition}},  {\em JHEP} {\bf 07} (2014) 108,
  [\href{http://arxiv.org/abs/1401.1827}{{\tt arXiv:1401.1827}}].

\bibitem{Guth:1981uk}
A.~H. Guth and E.~J. Weinberg, {\it {Cosmological Consequences of a First Order
  Phase Transition in the SU(5) Grand Unified Model}},  {\em Phys. Rev. D} {\bf
  23} (1981) 876.

\bibitem{Blasi:2023rqi}
S.~Blasi, R.~Jinno, T.~Konstandin, H.~Rubira, and I.~Stomberg, {\it
  {Gravitational waves from defect-driven phase transitions: domain walls}},
  {\em JCAP} {\bf 10} (2023) 051, [\href{http://arxiv.org/abs/2302.06952}{{\tt
  arXiv:2302.06952}}].

\bibitem{Coleman:1973jx}
S.~R. Coleman and E.~J. Weinberg, {\it {Radiative Corrections as the Origin of
  Spontaneous Symmetry Breaking}},  {\em Phys. Rev. D} {\bf 7} (1973)
  1888--1910.

\bibitem{Larsson:1996sp}
S.~E. Larsson, S.~Sarkar, and P.~L. White, {\it {Evading the cosmological
  domain wall problem}},  {\em Phys. Rev. D} {\bf 55} (1997) 5129--5135,
  [\href{http://arxiv.org/abs/hep-ph/9608319}{{\tt hep-ph/9608319}}].

\bibitem{Casini:2001ai}
H.~Casini and S.~Sarkar, {\it {No cosmological domain wall problem for weakly
  coupled fields}},  {\em Phys. Rev. D} {\bf 65} (2002) 025002,
  [\href{http://arxiv.org/abs/hep-ph/0106272}{{\tt hep-ph/0106272}}].

\bibitem{Guada:2020xnz}
V.~Guada, M.~Nemev\v{s}ek, and M.~Pintar, {\it {FindBounce: Package for
  multi-field bounce actions}},  {\em Comput. Phys. Commun.} {\bf 256} (2020)
  107480, [\href{http://arxiv.org/abs/2002.00881}{{\tt arXiv:2002.00881}}].

\bibitem{Wainwright:2011kj}
C.~L. Wainwright, {\it {CosmoTransitions: Computing Cosmological Phase
  Transition Temperatures and Bubble Profiles with Multiple Fields}},  {\em
  Comput. Phys. Commun.} {\bf 183} (2012) 2006--2013,
  [\href{http://arxiv.org/abs/1109.4189}{{\tt arXiv:1109.4189}}].

\bibitem{AMBROSETTI1973349}
A.~Ambrosetti and P.~H. Rabinowitz, {\it Dual variational methods in critical
  point theory and applications},  {\em Journal of Functional Analysis} {\bf
  14} (1973), no.~4 349--381.

\bibitem{Guth:1982pn}
A.~H. Guth and E.~J. Weinberg, {\it {Could the Universe Have Recovered from a
  Slow First Order Phase Transition?}},  {\em Nucl. Phys. B} {\bf 212} (1983)
  321--364.

\bibitem{PhysRevD.46.2384}
M.~S. Turner, E.~J. Weinberg, and L.~M. Widrow, {\it Bubble nucleation in
  first-order inflation and other cosmological phase transitions},  {\em Phys.
  Rev. D} {\bf 46} (Sep, 1992) 2384--2403.

\bibitem{Ellis:2018mja}
J.~Ellis, M.~Lewicki, and J.~M. No, {\it {On the Maximal Strength of a
  First-Order Electroweak Phase Transition and its Gravitational Wave Signal}},
   {\em JCAP} {\bf 04} (2019) 003, [\href{http://arxiv.org/abs/1809.08242}{{\tt
  arXiv:1809.08242}}].

\bibitem{Ellis:2020nnr}
J.~Ellis, M.~Lewicki, and V.~Vaskonen, {\it {Updated predictions for
  gravitational waves produced in a strongly supercooled phase transition}},
  {\em JCAP} {\bf 11} (2020) 020, [\href{http://arxiv.org/abs/2007.15586}{{\tt
  arXiv:2007.15586}}].

\bibitem{Vilenkin:2000jqa}
A.~Vilenkin and E.~P.~S. Shellard, {\em {Cosmic Strings and Other Topological
  Defects}}.
\newblock Cambridge University Press, 7, 2000.

\bibitem{Arnold:1993wc}
P.~B. Arnold, {\it {One loop fluctuation - dissipation formula for bubble wall
  velocity}},  {\em Phys. Rev. D} {\bf 48} (1993) 1539--1545,
  [\href{http://arxiv.org/abs/hep-ph/9302258}{{\tt hep-ph/9302258}}].

\bibitem{Moore:1995ua}
G.~D. Moore and T.~Prokopec, {\it {Bubble wall velocity in a first order
  electroweak phase transition}},  {\em Phys. Rev. Lett.} {\bf 75} (1995)
  777--780, [\href{http://arxiv.org/abs/hep-ph/9503296}{{\tt hep-ph/9503296}}].

\bibitem{Moore:1995si}
G.~D. Moore and T.~Prokopec, {\it {How fast can the wall move? A Study of the
  electroweak phase transition dynamics}},  {\em Phys. Rev. D} {\bf 52} (1995)
  7182--7204, [\href{http://arxiv.org/abs/hep-ph/9506475}{{\tt
  hep-ph/9506475}}].

\bibitem{Bodeker:2009qy}
D.~Bodeker and G.~D. Moore, {\it {Can electroweak bubble walls run away?}},
  {\em JCAP} {\bf 05} (2009) 009, [\href{http://arxiv.org/abs/0903.4099}{{\tt
  arXiv:0903.4099}}].

\bibitem{Bodeker:2017cim}
D.~Bodeker and G.~D. Moore, {\it {Electroweak Bubble Wall Speed Limit}},  {\em
  JCAP} {\bf 05} (2017) 025, [\href{http://arxiv.org/abs/1703.08215}{{\tt
  arXiv:1703.08215}}].

\bibitem{Gouttenoire:2021kjv}
Y.~Gouttenoire, R.~Jinno, and F.~Sala, {\it {Friction pressure on relativistic
  bubble walls}},  {\em JHEP} {\bf 05} (2022) 004,
  [\href{http://arxiv.org/abs/2112.07686}{{\tt arXiv:2112.07686}}].

\bibitem{Azatov:2023xem}
A.~Azatov, G.~Barni, R.~Petrossian-Byrne, and M.~Vanvlasselaer, {\it
  {Quantisation Across Bubble Walls and Friction}},
  \href{http://arxiv.org/abs/2310.06972}{{\tt arXiv:2310.06972}}.

\bibitem{Hindmarsh:2013xza}
M.~Hindmarsh, S.~J. Huber, K.~Rummukainen, and D.~J. Weir, {\it {Gravitational
  waves from the sound of a first order phase transition}},  {\em Phys. Rev.
  Lett.} {\bf 112} (2014) 041301, [\href{http://arxiv.org/abs/1304.2433}{{\tt
  arXiv:1304.2433}}].

\bibitem{Giblin:2014qia}
J.~T. Giblin and J.~B. Mertens, {\it {Gravitional radiation from first-order
  phase transitions in the presence of a fluid}},  {\em Phys. Rev. D} {\bf 90}
  (2014), no.~2 023532, [\href{http://arxiv.org/abs/1405.4005}{{\tt
  arXiv:1405.4005}}].

\bibitem{Hindmarsh:2015qta}
M.~Hindmarsh, S.~J. Huber, K.~Rummukainen, and D.~J. Weir, {\it {Numerical
  simulations of acoustically generated gravitational waves at a first order
  phase transition}},  {\em Phys. Rev. D} {\bf 92} (2015), no.~12 123009,
  [\href{http://arxiv.org/abs/1504.03291}{{\tt arXiv:1504.03291}}].

\bibitem{Hindmarsh:2016lnk}
M.~Hindmarsh, {\it {Sound shell model for acoustic gravitational wave
  production at a first-order phase transition in the early Universe}},  {\em
  Phys. Rev. Lett.} {\bf 120} (2018), no.~7 071301,
  [\href{http://arxiv.org/abs/1608.04735}{{\tt arXiv:1608.04735}}].

\bibitem{Hindmarsh:2017gnf}
M.~Hindmarsh, S.~J. Huber, K.~Rummukainen, and D.~J. Weir, {\it {Shape of the
  acoustic gravitational wave power spectrum from a first order phase
  transition}},  {\em Phys. Rev. D} {\bf 96} (2017), no.~10 103520,
  [\href{http://arxiv.org/abs/1704.05871}{{\tt arXiv:1704.05871}}]. [Erratum:
  Phys.Rev.D 101, 089902 (2020)].

\bibitem{Cutting:2018tjt}
D.~Cutting, M.~Hindmarsh, and D.~J. Weir, {\it {Gravitational waves from vacuum
  first-order phase transitions: from the envelope to the lattice}},  {\em
  Phys. Rev. D} {\bf 97} (2018), no.~12 123513,
  [\href{http://arxiv.org/abs/1802.05712}{{\tt arXiv:1802.05712}}].

\bibitem{Hindmarsh:2019phv}
M.~Hindmarsh and M.~Hijazi, {\it {Gravitational waves from first order
  cosmological phase transitions in the Sound Shell Model}},  {\em JCAP} {\bf
  12} (2019) 062, [\href{http://arxiv.org/abs/1909.10040}{{\tt
  arXiv:1909.10040}}].

\bibitem{Lewicki:2021pgr}
M.~Lewicki, M.~Merchand, and M.~Zych, {\it {Electroweak bubble wall expansion:
  gravitational waves and baryogenesis in Standard Model-like thermal plasma}},
   {\em JHEP} {\bf 02} (2022) 017, [\href{http://arxiv.org/abs/2111.02393}{{\tt
  arXiv:2111.02393}}].

\bibitem{Espinosa:2010hh}
J.~R. Espinosa, T.~Konstandin, J.~M. No, and G.~Servant, {\it {Energy Budget of
  Cosmological First-order Phase Transitions}},  {\em JCAP} {\bf 06} (2010)
  028, [\href{http://arxiv.org/abs/1004.4187}{{\tt arXiv:1004.4187}}].

\bibitem{Steinhardt:1981ct}
P.~J. Steinhardt, {\it {Relativistic Detonation Waves and Bubble Growth in
  False Vacuum Decay}},  {\em Phys. Rev. D} {\bf 25} (1982) 2074.

\bibitem{Adams:1989su}
F.~C. Adams, K.~Freese, and L.~M. Widrow, {\it {Evolution of Nonspherical
  Bubbles}},  {\em Phys. Rev. D} {\bf 41} (1990) 347.

\bibitem{Garriga:1991ts}
J.~Garriga and A.~Vilenkin, {\it {Perturbations on domain walls and strings: A
  Covariant theory}},  {\em Phys. Rev. D} {\bf 44} (1991) 1007--1014.

\bibitem{Garriga:1991tb}
J.~Garriga and A.~Vilenkin, {\it {Quantum fluctuations on domain walls, strings
  and vacuum bubbles}},  {\em Phys. Rev. D} {\bf 45} (1992) 3469--3486.

\bibitem{Thrane:2013oya}
E.~Thrane and J.~D. Romano, {\it {Sensitivity curves for searches for
  gravitational-wave backgrounds}},  {\em Phys. Rev. D} {\bf 88} (2013), no.~12
  124032, [\href{http://arxiv.org/abs/1310.5300}{{\tt arXiv:1310.5300}}].

\bibitem{Sesana:2019vho}
A.~Sesana et~al., {\it {Unveiling the gravitational universe at $\mu$-Hz
  frequencies}},  {\em Exper. Astron.} {\bf 51} (2021), no.~3 1333--1383,
  [\href{http://arxiv.org/abs/1908.11391}{{\tt arXiv:1908.11391}}].

\bibitem{Anderson:1991zb}
G.~W. Anderson and L.~J. Hall, {\it {The Electroweak phase transition and
  baryogenesis}},  {\em Phys. Rev. D} {\bf 45} (1992) 2685--2698.

\end{thebibliography}\endgroup


\end{document}